\documentclass[10pt]{article}

\usepackage{graphicx}
\usepackage{amsmath}
\usepackage{amssymb}
\usepackage[cp1251]{inputenc}
\usepackage[T2A]{fontenc}
\numberwithin{equation}{section}
\newcommand{\eqa}{\begin{eqnarray}}
\newcommand{\eeqa}{\end{eqnarray}}
\newcommand{\beq}{\begin{equation}}
\newcommand{\eeq}{\end{equation}}

\begin{document}
\parskip 6pt
\hoffset -1.8cm

\title{Integrable Zhaidary   equations:   reductions and gauge equivalence}
\author{Zh. Sagidullayeva$^{1}$\footnote{Email: zrmyrzakulova@gmail.com},  \,    K. Yesmakhanova$^{1}$\footnote{Email: krmyrzakulova@gmail.com},     \,   R. Myrzakulov$^{1}$\footnote{Email: rmyrzakulov@gmail.com}, \, Zh. Myrzakulova$^{1}$\footnote{Email: zhrmyrzakulova@gmail.com}, \\ N.  Serikbayev$^{1}$\footnote{Email: ns.serikbayev@gmail.com}, \,   
G.   Nugmanova$^{1}$\footnote{Email: nugmanovagn@gmail.com}, 
\,  A. Sergazina$^{1}$\footnote{Email: asergazina@gmail.com} \,  and \,
K. Yerzhanov$^{1}$\footnote{Email:yerzhanovkk@gmail.com} \\
\textsl{$^{1}$Ratbay Myrzakulov Eurasian International Centre for Theoretical Physics}, \\ \textsl{Astana, 010009, Kazakhstan}\\   
}
\date{}
\maketitle

\begin{abstract} The present work addresses the study and characterization of the integrability of some  spin systems (ISS)  in 1+1 dimensions. Lax representations  for these ISS  are successfully obtained. The gauge equivalent counterparts of these integrable ISS are presented. Finally, we consider Zhanbota transcendents and  some integrable Zhanbota equations. In particular, the gauge equivalence between some  Zhanbota equations and the six Painleve equations is established. 

\end{abstract}

{\bf Key words}: Integrable equations, nonlinear Schr\"{o}dinger equation, Heisenberg ferromagnet equation,  Landau-Lifshitz equation, soliton equations, gauge equivalence, Zhanbota transcendents, derivative nonlinear Schr\"{o}dinger equation,   Zhanbota equations,  spin systems, Painleve equations.

\tableofcontents

\section{Introduction}
The Heisenberg ferromagnet equation (HFE)
\begin{eqnarray}
iS_{t}+\frac{1}{2}[S,S_{xx}]=0
\end{eqnarray}
is  one of the most fundamental  integrable equations in soliton theory. It is well known that some  integrable nonlinear evolution equations are related, often in an unexpected way, to each other by means of so-called gauge equivalences. For example, the gauge partner of the HFE   is the following nonlinear Schr\"{o}dinger  equation (NLSE) \cite{laksh}-\cite{zt1979}
\begin{eqnarray}
iq_{t}+q_{xx}+2\epsilon|q|^{2}q=0.
\end{eqnarray}
Gauge equivalences between integrable nonlinear equations can be exploited to gain insight into
either equation  from its gauge partner, for instance by transforming solutions of one equation to solutions of the other. 
Formally the system can be cast into two gauge equivalent zero curvature conditions for
the two sets of  Lax operators. Returning to the HFE (1.1)  and NLSE (1.2) we can note that they  play important role in modern nonlinear physics and mathematics. They  admit several integrable generalizations in 1+1 and 2+1 dimensions (see, e.g. \cite{ishimori}-\cite{s4} and references therein). In this paper, for example among the  integrable
generalizations of HFE and NLSE, we are interested in the study of their  derivative versions. In particular, for the NLSE (1.2) there exist the following three celebrated derivative nonlinear Schr\"{o}dinger  equations (DNLSE) of this
kind, the DNLSE-I, DNLSE-II and the DNLSE-III (see, e.g. \cite{2102.12183} and references therein):

i) the Kaup-Newell equation (KNE) or DNLSE-I \cite{kne}
\begin{eqnarray}
iq_{t}-q_{xx}-i(|q|^{2}q)_{x}=0,
\end{eqnarray}

ii) the Chen-Lee-Liu equation (CLLE) or DNLSE-II \cite{clle}
 \begin{eqnarray}
iq_{t}-q_{xx}-i|q|^{2}q_{x}=0,
\end{eqnarray}

iii) the Gerdjikov-Ivanov equation (GIE) or DNLSE-III \cite{gie}
\begin{eqnarray}
iq_{t}-q_{xx}+iq^{2}\bar{q}_{x}-0.5|q|^{4}q=0.
\end{eqnarray}
Through the $U(1)$ - gauge transformation, these  three  DNLSE  can be related to each other.  In fact, if $q(x,t)$ 
 is a solution of the KNE (1.3), it is easy to find that the
new field $p(x, t)$ defined as (see, e.g. \cite{2102.12183} and references therein)
\begin{eqnarray}
p(x,t)=q(x,t)e^{0.5\delta \zeta(x,t)},
\end{eqnarray}
with 
\begin{eqnarray}
\zeta_{x}=|q|^{2}, \quad \zeta_{t}=i(q\bar{q}_{x}-
\bar{q}q_{x})+1.5|q|^{4}, \quad \delta=const
\end{eqnarray}
satisfies the CLLE (1.4) for $\delta = 1$, and the GIE 
 (1.5) for $\delta=2$.

In the modern mathematics and physics, gauge transformations constitute an useful tool to link integrable nonlinear evolution  equations (INLEE) 
in soliton theory, since they provide some  transformations (like Darboux,
  B\"{a}cklund) between those INLEE as well as the relation of their associated linear problems (Lax representations) (see, e.g. \cite{1612.06723}-\cite{rm21} and references therein). In this paper, 
we exploit this gauge transformation property to construct the gauge equivalent counterpart of the  Zhaidary  equation and to find its integrable reductions and generalizations. In particular, the gauge equivalence between some  Zhanbota equations and the six Painleve equations is established.

This  paper is organized as follows.  In section 2 we study the Zhaidary equation (ZE).  In section 3,  we construct the gauge equivalent partner of the ZE.  The section 4, provides the particular reductions of the ZE and their gauge
equivalents and gives the associated new Lax operators in the
explicit form. Some integrable generalizations of the ZE is presented in section 5. Finally in Section 6, we consider Zhanbota transcendents and  integrable Zhanbota equations. Our conclusions are stated in section 7.

%%%%%%%%%%%%%%%%%%%%%%%%%%%%%%%%%%%%%%%%%%%%%%%%%%%%%
\section{Zhaidary  equation}
%%%%%%%%%%%%%%%%%%%%%%%%%%%%%%%%%%%%%%%%%%%%%%%%%%%%%%%
The anisotropic Zhaidary-I equation (Z-IE) is given by \cite{2205.02073}-\cite{z3}, \cite{49}
\begin{eqnarray}
(1+2\beta(c\beta+d)){\bf S}_{t}-{\bf S}\wedge {\bf S}_{xt}-u{\bf S}_{x}+4cw{\bf S}_{x}-{\bf S}\wedge J{\bf S}&=&0,\\
u_{x}+\frac{1}{2}({\bf S}_{x}^{2})_{t}&=&0,\\
w_{x}+\frac{1}{4(2\beta c+d)^{2}}({\bf S}_{x}^{2})_{t}&=&0,
\end{eqnarray}
where ${\bf S}=(S_{1},S_{2},S_{3})$ is the unit spin vector, ${\bf S}^{2}=1$, $J=diag(J_{1}, J_{2}, J_{3})$ ($J_{1}\leq J_{2} \leq J_{3}$), $u$ and $v$ are  scalar functions (potentials). The isotropic Z-IE has the form
\begin{eqnarray}
(1+2\beta(c\beta+d)){\bf S}_{t}-{\bf S}\wedge {\bf S}_{xt}-u{\bf S}_{x}-2\beta(c\beta+d){\bf S}_{t}+4cw{\bf S}_{x}&=&0,\\
u_{x}+\frac{1}{2}({\bf S}_{x}^{2})_{t}&=&0,\\
w_{x}+\frac{1}{4(2\beta c+d)^{2}}({\bf S}_{x}^{2})_{t}&=&0.
\end{eqnarray}
The Z-IE (2.4)-(2.6) we simetime  call the spin Zhaidary equation. The isotropic Z-IE is  completely integrable that can be solved by the inverse scattering transformation method (IST). It possesses all the
basic characters of integrable equations.  The corresponding Lax representation (LR)  has the form
\begin{eqnarray}
\Psi_{x}&=&U_{1}\Psi, \\ 
\Psi_{t}&=&V_{1}\Psi. 
\end{eqnarray}
Here
\begin{eqnarray}
U_{1}&=&[ic(\lambda^{2}-\beta^{2})+id(\lambda-\beta)] S+\frac{c(\lambda-\beta)}{2c\beta+d}SS_{x}, \\
V_{1}&=&\frac{1}{1-2c\lambda^{2}-2d\lambda}\{[2c(\lambda^{2}-\beta^{2})+2d(\lambda-\beta)]B+\lambda^{2}F_{2}+\lambda F_{1}+F_{0}\}, 
\end{eqnarray}
where 
\begin{eqnarray}
F_{2}&=&-4ic^{2}wS, \quad F_{1}=-4icdwS-\frac{4c^{2}}{2c\beta+d}wSS_{x}-\frac{ic}{2c\beta+d}S\{(SS_{x})_{t}-[SS_{x},B]\}, \\
F_{0}&=&-\beta F_{1}-\beta^{2}F_{2}, \quad B=0.25([S,S_{t}]+2iuS), \quad S={\bf S}\cdot {\bf \sigma}.
\end{eqnarray}
If $\beta=0$, the Z-IE takes the form
\begin{eqnarray}
{\bf S}_{t}-{\bf S}\wedge {\bf S}_{xt}-u{\bf S}_{x}+4cw{\bf S}_{x}&=&0,\\
u_{x}+\frac{1}{2}({\bf S}_{x}^{2})_{t}&=&0,\\
w_{x}+\frac{1}{4d^{2}}({\bf S}_{x}^{2})_{t}&=&0,
\end{eqnarray}
or
\begin{eqnarray}
{\bf S}_{t}-{\bf S}\wedge {\bf S}_{xt}+(2cd^{-2}-1)u{\bf S}_{x}&=&0,\\
u_{x}+\frac{1}{2}({\bf S}_{x}^{2})_{t}&=&0.
\end{eqnarray}

%%%%%%%%%%%%%%%%%%%%%%%%%%%%%%%%%%%%
\section{Gauge  equivalent counterpart}
%%%%%%%%%%%%%%%%%%%%%%%%%%%%%%%%%%%%%%

Let us find the gauge equivalent counterpart of the Z-IE (2.4)-(2.6). For this aim, consider the gauge transformation
\begin{eqnarray}
\Phi=g \Psi, \quad g=\Phi|_{\lambda=\beta}, \quad S=g^{-1}\sigma_{3}g.
\end{eqnarray}
Then the new function $\Phi$ satisfies the following set of the linear equations
\begin{eqnarray}
\Phi_{x}&=&U_{2}\Phi, \\ 
\Phi_{t}&=&V_{2}\Phi, 
\end{eqnarray}
where
\begin{eqnarray}
U_{2}&=&[i(c\lambda^{2}+d\lambda)\sigma_{3}+(2c\lambda+d)Q, \\
V_{2}&=&\frac{1}{1-2c\lambda^{2}-2d\lambda}(\lambda^{2}B_{2}+\lambda B_{1}+B_{0}). 
\end{eqnarray}
Here
\begin{eqnarray}
B_{2}=-4ic\sigma_{3}, \quad B_{1}=-4icdv\sigma_{3}-2ic\sigma_{3}Q_{t}-8c^{2}vQ, \quad B_{0}=\frac{d}{2c}B_{1}-\frac{d^{2}}{4c^{2}}B_{2}, 
\end{eqnarray}
and
\begin{eqnarray}
Q=\left(\begin{array}{cc} 0 & q \\ r & 0  \end{array}\right), \quad r=\epsilon \bar{q}, \quad \epsilon=\pm 1.
\end{eqnarray}
The compatibility condition 
\begin{eqnarray}
U_{2t}-V_{2x}+[U_{2},V_{2}]=0
\end{eqnarray}
gives the following equation
\begin{eqnarray}
iq_{t}-q_{xt}+4ic(vq)_{x}-2d^{2}vq&=&0,\\ 
ir_{t}+r_{xt}+4ic(vr)_{x}+2d^{2}vr&=&0,\\
v_{x}-(rq)_{t}&=&0,
\end{eqnarray}
which is  called the  Zhaidary-II equation (Z-IIE).
%%%%%%%%%%%%%%%%%%%%%%%%%%%%%%%%%%%%%%%%%%%%%%%%%%%%
\section{Integrable reductions}
%%%%%%%%%%%%%%%%%%%%%%%%%%%%%%%%%%%%%%%%%%%%%%%%
The Zhaidary equations (ZE) admit several integrable reductions. Let us present these particular cases.
%%%%%%%%%%%%%%%%%%%%%%%%%%%%%%%%%%
\subsection{Kuralay equation}
%%%%%%%%%%%%%%%%%%%%%%%%%%%%%%%%%%%%
First, we assume that $c=0$. Then from the Z-IE (2.4)-(2.6) we obtain the following  Kuralay-I equation (K-IE) \cite{2206.05348}-\cite{Zafar}, \cite{47}:
\begin{eqnarray}
(1+2\beta d){\bf S}_{t}-{\bf S}\wedge {\bf S}_{xt}-u{\bf S}_{x}&=&0,\\
u_{x}+\frac{1}{2}({\bf S}_{x}^{2})_{t}&=&0.
\end{eqnarray}
 If $\beta=0$, this K-IE takes the form
\begin{eqnarray}
{\bf S}_{t}-{\bf S}\wedge {\bf S}_{xt}-u{\bf S}_{x}&=&0,\\
u_{x}+\frac{1}{2}({\bf S}_{x}^{2})_{t}&=&0.
\end{eqnarray}
The  K-IE is integrable by IST method. The corresponding LR looks like 
\begin{eqnarray}
\Psi_{x}&=&U_{3}\Psi, \\ 
\Psi_{t}&=&V_{3}\Psi. 
\end{eqnarray}
Here
\begin{eqnarray}
U_{3}&=&id(\lambda-\beta)S, \quad S=\left(\begin{array}{cc} S_{3} & S^{-}\\ S^{+} & -S_{3}  \end{array}\right), \quad S^{2}=I, \quad \quad S^{\pm}=S_{1}\pm iS_{2},\\
V_{3}&=&\frac{1}{1-2d\lambda}\{2d(\lambda-\beta)]B+\lambda^{2}F_{2}+\lambda F_{1}+F_{0}\}, 
\end{eqnarray}
where 
\begin{eqnarray}
F_{2}&=&0, \quad F_{1}=0, \quad F_{0}=0, \quad B=0.25([S,S_{t}]+2iuS)=0.25 Z. 
\end{eqnarray}

Let us  now we present  the gauge equivalent counterpart of the K-IE (4.3)-(4.4). As usual, we  consider the gauge transformation
\begin{eqnarray}
\Phi=g \Psi, \quad g=\Phi|_{\lambda=\beta}, \quad S=g^{-1}\sigma_{3}g.
\end{eqnarray}
Then the new function $\Phi$ satisfies the following set of the linear equations
\begin{eqnarray}
\Phi_{x}&=&U_{4}\Phi, \\ 
\Phi_{t}&=&V_{4}\Phi, 
\end{eqnarray}
where
\begin{eqnarray}
U_{4}&=&[id\lambda\sigma_{3}+dQ, \\
V_{4}&=&\frac{1}{1-2d\lambda}(\lambda^{2}B_{2}+\lambda B_{1}+B_{0}). 
\end{eqnarray} 
Here
\begin{eqnarray}
B_{2}=0, \quad B_{1}=0, \quad B_{0}=\frac{d}{2}[-4idv\sigma_{3}-2i\sigma_{3}Q_{t}-8cvQ,], 
\end{eqnarray}
and
\begin{eqnarray}
Q=\left(\begin{array}{cc} 0 & q \\ r & 0  \end{array}\right).
\end{eqnarray}
The compatibility condition 
\begin{eqnarray}
U_{4t}-V_{4x}+[U_{4},V_{4}]=0
\end{eqnarray}
is equivalent to  the following Kuralay-II equation (K-IIE):
\begin{eqnarray}
iq_{t}-q_{xt}-2d^{2}vq&=&0,\\ 
ir_{t}+r_{xt}+2d^{2}vr&=&0,\\
v_{x}-(rq)_{t}&=&0. 
\end{eqnarray}

\subsection{Shynaray equation}
Our next particular case is when $d=0$. Then from the Z-IE (2.1)-(2.3) we get  the following  Shynaray-I equation (S-IE):
\begin{eqnarray}
(1+2c\beta^{2}){\bf S}_{t}-{\bf S}\wedge {\bf S}_{xt}-u{\bf S}_{x}+4cw{\bf S}_{x}&=&0,\\
u_{x}+\frac{1}{2}({\bf S}_{x}^{2})_{t}&=&0,\\
w_{x}+\frac{1}{16c^{2}\beta^{2}}({\bf S}_{x}^{2})_{t}&=&0,
\end{eqnarray}
As the particular case of the integrable equation, the  S-IE is also integrable by the IST method. The corresponding LR is given by
\begin{eqnarray}
\Psi_{x}&=&U_{5}\Psi, \\ 
\Psi_{t}&=&V_{5}\Psi. 
\end{eqnarray}
Here
\begin{eqnarray}
U_{5}&=&ic(\lambda^{2}-\beta^{2}) S+\frac{\lambda-\beta}{2\beta}SS_{x}, \\
V_{5}&=&\frac{1}{1-2c\lambda^{2}}\{2c(\lambda^{2}-\beta^{2})B+\lambda^{2}F_{2}+\lambda F_{1}+F_{0}\}, 
\end{eqnarray}
where 
\begin{eqnarray}
F_{2}&=&-4ic^{2}wS, \quad F_{1}=-\frac{4c^{2}}{2c\beta}wSS_{x}-\frac{ic}{2c\beta}S\{(SS_{x})_{t}-[SS_{x},B]\}, \\
F_{0}&=&-\beta F_{1}-\beta^{2}F_{2}, \quad B=0.25([S,S_{t}]+2iuS), \quad S={\bf S}\cdot {\bf \sigma}.
\end{eqnarray}

Let us find the gauge equivalent counterpart of the S-IE (4.21)-(4.23). Consider the gauge transformation
\begin{eqnarray}
\Phi=g \Psi, \quad g=\Phi|_{\lambda=\beta}, \quad S=g^{-1}\sigma_{3}g.
\end{eqnarray}
Then the  function $\Phi$ satisfies the following set of the linear equations
\begin{eqnarray}
\Phi_{x}&=&U_{6}\Phi, \\ 
\Phi_{t}&=&V_{6}\Phi, 
\end{eqnarray}
where
\begin{eqnarray}
U_{6}&=&ic\lambda^{2}\sigma_{3}+2c\lambda Q, \\
V_{6}&=&\frac{1}{1-2c\lambda^{2}}(\lambda^{2}B_{2}+\lambda B_{1}+B_{0}). 
\end{eqnarray}
Here
\begin{eqnarray}
B_{2}=-4ic\sigma_{3}, \quad B_{1}=-2ic\sigma_{3}Q_{t}-8c^{2}vQ, \quad B_{0}=0. 
\end{eqnarray}

Hence, the compatibility condition of the equations (4.31)-(4.32) 
\begin{eqnarray}
U_{6t}-V_{6x}+[U_{6},V_{6}]=0
\end{eqnarray}
gives the following nonlinear evolution equation
\begin{eqnarray}
iq_{t}-q_{xt}+4ic(vq)_{x}&=&0,\\ 
ir_{t}+r_{xt}+4ic(vr)_{x}&=&0,\\
v_{x}-(rq)_{t}&=&0. 
\end{eqnarray}
It is the Shynaray-II equation (S-IIE).
%%%%%%%%%%%%%%%%%%%%%%%%%%%%%%%%%%%%%%
\section{Integrable generalizations}
%%%%%%%%%%%%%%%%%%%%%%%%%%%%%%%%%%%%%
In the previous section we have presented some integrable reductions of the ZE (2.4)-(2.6). It is interesting to note that this  ZE also admits several integrable extensions/generalizations. Now we are going to present some of these integrable generalizations of the ZE.

%%%%%%%%%%%%%%%%%%%%%%%%%%%%%%%%%%%%%%%%%%%%%%%%%%%%%%%
\subsection{Nurshuak equation}
%%%%%%%%%%%%%%%%%%%%%%%%%%%%%%%%%%%%%%%%%%%%%%%%%%%

 One of such integrable generalizations of the ZE (2.4)-(2.6) is the following Nurshuak-I equation (N-IE): 
\begin{eqnarray}
iS_{t}+2\epsilon_{1}Z_x+i\epsilon_{2}(S_{xt}+[S_x,Z])_{x}+(wS)_{x}+\frac{1}{\omega}[S, W]&=&0\label{5.1},\\
u_x-\frac{i}{4}tr(S\times[S_x,S_t])&=&0\label{5.2},\\
w_x-\frac{i}{4}\epsilon_{2}[tr(S_x^2)]_{t}&=&0\label{5.3},\\
 iW_{x}+\omega [S, W]&=&0\label{5.4},
\end{eqnarray} 
 where
\begin{eqnarray}
Z=\frac{1}{4}([S,S_{t}]+2iuS). 
\end{eqnarray}

As integrable equation, the N-IE  admits the following  LR:
 \begin{eqnarray}
\Psi_{x}&=&U_{7}\Psi, \label{5.6}\\
\Psi_{t}&=&V_{7}\Psi. \label{5.7} 
\end{eqnarray} 
Here
 \begin{eqnarray}
U_{7}&=&-i\lambda S\label{5.8},\\
V_{7}&=&\frac{1}{1-2\epsilon_{1}\lambda-4\epsilon_{2}\lambda^2}\{(2\epsilon_{1}\lambda+4\epsilon_{2}\lambda^{2})Z+\lambda V_{1}+\frac{i}{\lambda+\omega}W-\frac{i}{\omega}W\}, \label{5.9} 
\end{eqnarray} 
where
\begin{eqnarray}
V_1&=&wS+i\epsilon_2(S_{xt}+[S_x,Z]), \label{5.10}\\
W&=&\begin{pmatrix} W_3&W^{-}\\W^{+}& -W_3\end{pmatrix}, \quad S=\begin{pmatrix} S_3&S^{-}\\S^{+}& -S_3\end{pmatrix}, \quad S^{\pm}=S_{1}\pm i S_{2}. \label{5.11} 
\end{eqnarray} 
 The  compatibility condition $\Psi_{xt}=\Psi_{tx}$ gives the N-IE (\ref{5.1})--(\ref{5.4}). The  gauge partner  of the N-IE (\ref{5.1})--(\ref{5.4})  is the following Nurshuak-II equation (N-IIE)
 \begin{eqnarray}
iq_{t}+\epsilon_1q_{xt}+i\epsilon_2q_{xxt}-vq+(wq)_x-2ip&=&0 \label{5.26},\\
ir_{t}-\epsilon_1r_{xt}+i\epsilon_2r_{xxt}+vr+(wr)_x-2ik&=&0\label{5.27},\\
v_{x}+2\epsilon_1(rq)_{t}-2i\epsilon_2(r_{xt}q-rq_{xt})&=&0\label{5.28},\\
w_{x}-2i\epsilon_2(rq)_{t}&=&0\label{5.29},\\
p_{x}-2i\omega p -2\eta q&=&0\label{5.30},\\
k_x+2i\omega k-2\eta r&=&0\label{5.31},\\
\eta_{x}+r p +k q&=&0\label{5.32},
\end{eqnarray}
where $r=\delta_{1}\bar{q}, \quad k=\delta_{2}\bar{p}, \quad \delta_{j}=\pm 1$.  As the gauge equivalent of the integrable N-IE, this N-IIE (\ref{5.26}) - (\ref{5.32}) is also integrable. Its  LR  reads as
\begin{eqnarray}
\Psi_{x}&=&U_{8}\Psi\label{5.33},\\
\Psi_{t}&=&V_{8}\Psi\label{5.34}.
\end{eqnarray} 
Here
 \begin{eqnarray}
U_{8}&=&-i\lambda \sigma_3+A_0\label{5.35},\\
V_{8}&=&\frac{1}{1-(2\epsilon_1\lambda+4\epsilon_2\lambda^2)}\{\lambda B_1+B_0+\frac{i}{\lambda+\omega}B_{-1}\}\label{5.36},
\end{eqnarray} 
where
\begin{eqnarray}
B_1&=&w\sigma_3+2i\epsilon_2\sigma_3A_{0t}, \quad A_0=\begin{pmatrix} 0&q\\-r& 0\end{pmatrix}\label{5.38},\\
B_0&=&-\frac{i}{2}v\sigma_3+\begin{pmatrix} 0&i\epsilon_1q_t-\epsilon_2q_{xt}+iwq\\i\epsilon_1r_t+\epsilon_2r_{xt}-iwr& 0\end{pmatrix}\label{5.39},\\
B_{-1}&=&\begin{pmatrix} \eta&-p\\-k& -\eta\end{pmatrix}\label{5.40}. 
\end{eqnarray}

%%%%%%%%%%%%%%%%%%%%%%%%%%%%%%%%%%%%
\subsection{Aizhan equation}
%%%%%%%%%%%%%%%%%%%%%%%%%%%%%%%%%%%
The  Aizhan-I equation  is given by
\begin{eqnarray}
iS_{t}+i\epsilon_2(S_{xt}+[S_x,Z])_{x}+(wS)_{x}+\frac{1}{\omega}[S, W]&=&0\label{4.1},\\
u_x-\frac{i}{4}tr(S\times[S_x,S_t])&=&0\label{4.2},\\
w_x-\frac{i}{4}\epsilon_2[tr(S_x^2)]_t&=&0\label{4.3},\\
iW_{x}+\omega [S, W]&=&0\label{4.4}.
\end{eqnarray}
The Aizhan-I equation is one of integrable generalizations of the Z-IE. The  Lax representation of the Aizhan-I equation has the form
\begin{eqnarray}
\Psi_{x}&=&U_{9}\Psi\label{4.6},\\
\Psi_{t}&=&V_{9}\Psi\label{4.7},
\end{eqnarray} 
where
\begin{eqnarray}
U_{9}&=&-i\lambda S\label{4.8},\\
V_{9}&=&\frac{1}{1-4\epsilon_2\lambda^2}\{4\epsilon_2\lambda^2Z+\lambda V_{1}+\frac{i}{\lambda+\omega}W-\frac{i}{\omega}W\}\label{4.9} 
\end{eqnarray} 
with
\begin{eqnarray}
V_1&=&wS+i\epsilon_2(S_{xt}+[S_x,Z])\label{4.10},\\
W&=&\begin{pmatrix} W_3&W^{-}\\W^{+}& -W_3\end{pmatrix}\label{4.11}. 
\end{eqnarray} 
The gauge equivalent counterpart of  the Aizhan-I equation (5.26)-(5.29) is the following Aizhan-II equation
 \begin{eqnarray}
iq_{t}+i\epsilon_2q_{xxt}-vq+(wq)_x-2ip&=&0 \label{4.19},\\
ir_{t}+i\epsilon_2r_{xxt}+vr+(wr)_x-2ik&=&0\label{4.20},\\
v_{x}-2i\epsilon_2(r_{xt}q-rq_{xt})&=&0\label{4.21},\\
w_{x}-2i\epsilon_2(rq)_t&=&0\label{4.22},\\
p_{x}-2i\omega p -2\eta q&=&0\label{4.23},\\
k_x+2i\omega k-2\eta r&=&0\label{4.24},\\
\eta_{x}+r p +k q&=&0\label{4.25}.
 \end{eqnarray}
 This  Aizhan-II  equation is integrable. The corresponding Lax representation is given by
\begin{eqnarray}
\Phi_{x}&=&U_{10}\Phi\label{4.26},\\
\Phi_{t}&=&V_{10}\Phi\label{4.27},
\end{eqnarray} 
where 
 \begin{eqnarray}
U_{10}&=&-i\lambda \sigma_3+A_0\label{4.28},\\
V_{10}&=&\frac{1}{1-4\epsilon_2\lambda^2}\{\lambda B_1+B_0+\frac{i}{\lambda+\omega}B_{-1}\}.\label{4.29} 
\end{eqnarray} 
Here
\begin{eqnarray}
B_1&=&w\sigma_3+2i\epsilon_2\sigma_3A_{0t}\label{4.30},\\
A_0&=&\begin{pmatrix} 0&q\\-r& 0\end{pmatrix}\label{4.31},\\
B_0&=&-\frac{i}{2}v\sigma_3+\begin{pmatrix} 0&-\epsilon_2q_{xt}+iwq\\\epsilon_2r_{xt}-iwr& 0\end{pmatrix}\label{4.32},\\
B_{-1}&=&\begin{pmatrix} \eta&-p\\-k& -\eta\end{pmatrix}\label{4.33}. 
\end{eqnarray}

%%%%%%%%%%%%%%%%%%%%%%%%%%%%%%%%%%%%
\subsection{Zhanbota equation}
%%%%%%%%%%%%%%%%%%%%%%%%%%%%%%%%%%%
In this section, we consider the Zhanbota-I equation. It has the form \cite{Guo1}
\begin{eqnarray}
{\bf S}_{t}-{\bf S}\wedge{\bf S}_{xt}-u{\bf S}_{x}-\frac{1}{\omega}{\bf S}\wedge {\bf W}&=&0\label{3.1},\\
u_x+{\bf S}\times(\bf S}_{x}\wedge{\bf S_{t})&=&0\label{3.2},\\
 {\bf W}_{x}-\omega {\bf S}\wedge{\bf W}&=&0. \label{3.3}
\end{eqnarray} 
The Zhanbota-I equation is also one of integrable generalization of the Zhaidary-I equation. The matrix form of the Zhanbota-I equation reads as
\begin{eqnarray}
iS_{t}+\frac{1}{2}[S, S_{xt}]+iuS_{x}+\frac{1}{\omega}[S, W]&=&0\label{3.4},\\
u_x-\frac{i}{4}tr(S[S_x,S_t])&=&0\label{3.5},\\
 iW_{x}+\omega [S, W]&=&0\label{3.6},
\end{eqnarray} 
where, $S=S_i\sigma_i$, $W=W_i\sigma_i$,($i=1,2,3$) and $\omega$ is a constant parameter. The  ${\bf W}=(W_1,W_2,W_3)$ is the vector potential. The Zhanbota-I equation  possesses the following  Lax representation:
 \begin{eqnarray}
\Psi_{x}&=&U_{11}\Psi\label{3.9},\\
\Psi_{t}&=&V_{11}\Phi\label{3.10}. 
\end{eqnarray} 
Here, the matrix operators $U_{11}$ and $V_{11}$ have the forms 
 \begin{eqnarray}
U_{11}&=&-i\lambda S\label{3.11},\\
V_{11}&=&\frac{1}{1-2\lambda}\{\lambda V_{1}+\frac{i}{\lambda+\omega}W-\frac{i}{\omega}W\}, \label{3.12}
\end{eqnarray} 
where
\begin{eqnarray}
V_1&=&2Z=\frac{1}{2}([S, S_{t}]+2iuS)\label{3.13},\\
W&=&\begin{pmatrix} W_3&W^{-}\\W^{+}& -W_3\end{pmatrix}\label{3.14}. 
\end{eqnarray} 

 Let us find the gauge equivalent counterpart of the Zhanbota-I equation (\ref{3.1}) - (\ref{3.3}). It is not difficult  to verify that the gauge equivalent counterpart of the Zhanbota-I  equation is the following Zhanbota-II equation:
\begin{eqnarray}
q_{t}+\frac{\kappa}{2i}q_{xt}+ivq-2p&=&0 \label{3.15},\\
r_{t}-\frac{\kappa}{2i}r_{xt}-ivr-2k&=&0\label{3.16},\\
v_{x}+\frac{\kappa}{2}(rq)_{t}&=&0\label{3.17},\\
p_{x}-2i\omega p -2\eta q&=&0\label{3.18},\\
k_x+2i\omega k-2\eta r&=&0\label{3.19},\\
\eta_{x}+r p +k q&=&0\label{3.20},
 \end{eqnarray}
where $q,r,p, k$ are some complex functions; $v, \eta$ are real potential functions and $\kappa$ is a constant parameter. The  Zhanbota-II equation is integrable. The  Lax representation of the Zhanbota-II  equations (\ref{3.15}) - (\ref{3.20}) reads as
\begin{eqnarray}
\Phi_{x}&=&U_{12}\Phi\label{3.21},\\
\Phi_{t}&=&V_{12}\Phi\label{3.22}, 
\end{eqnarray} 
where 
\begin{eqnarray}
U_{12}&=&-i\lambda \sigma_3+A_0\label{3.23},\\
V_{12}&=&\frac{1}{1-\kappa\lambda}\{B_0+\frac{i}{\lambda+\omega}B_{-1}\}.\label{3.24}
\end{eqnarray} 
Here
\begin{eqnarray}
A_0&=&\begin{pmatrix} 0&q\\-r& 0\end{pmatrix}\label{3.25},\\
B_0&=&-\frac{i}{2}v\sigma_3-\frac{\kappa}{2i}\begin{pmatrix} 0&q_y\\r_y& 0\end{pmatrix}\label{3.26},\\
B_{-1}&=&\begin{pmatrix} \eta&-p\\-k& -\eta\end{pmatrix}\label{3.27}. 
\end{eqnarray}
Next, we consider the reduction $r=\delta_{1} \bar{q}, \quad k=\delta_{2}\bar{p}$ with $\kappa=2, \quad \delta_{j}=\pm 1$, where the bar  means the complex conjugate. Then, the Zhanbota-II equation  (\ref{3.15})--(\ref{3.20}) takes the following more compact form
 \begin{eqnarray}
iq_{t}+q_{xt}-vq-2ip&=&0 \label{3.45},\\
v_{x}+2\delta_{1}(|q|^2)_{t}&=&0\label{3.46},\\
p_{x}-2i\omega p -2\eta q&=&0\label{3.47},\\
\eta_{x}+(\delta_{1}\bar{q} p +\delta_{2}\bar{p} q)&=&0. \label{3.48}
\end{eqnarray}
Note that it is nothing but the  Zhanbota-II equation. 
%%%%%%%%%%%%%%%%%%%%%%%%%%%%%%%%%%%%
\subsection{Akbota  equation}
%%%%%%%%%%%%%%%%%%%%%%%%%%%%%%%%%%%
In this subsection, we consider the following Akbota-I equation \cite{Guo2}-\cite{Thilagarajah}, \cite{47}
 \begin{eqnarray}
{\bf S}_{t}-{\bf S}\wedge (\alpha S_{xx}+\beta{\bf S}_{xt})-u{\bf S}_{x}&=&0\label{2.23},\\
 u_x+{\bf S}\times({\bf S}_{x}\wedge {\bf S}_{t})&=&0\label{2.24}. \end{eqnarray} 
 This Akbota-I equation is one of the integrable generalizations of the Zhaidary-I equation. It has the following Lax representation
\begin{eqnarray}
\Psi_{x}-U_{13}\Psi&=&0\label{2.25},\\
\Psi_{t}-V_{13}\Psi&=&0\label{2.26}, 
\end{eqnarray} where 
 \begin{equation}
U_{13}=\frac{i}{2}\lambda S, \quad V_{13}=\frac{1}{1-\lambda\beta}\{\alpha(\frac{1}{2}i\lambda^2 S+\frac{1}{4}[S,S_x])+\beta\lambda Z\}.\label{2.27}
\end{equation} 
 Note  that the gauge equivalent counterpart of the Akbota-I equation   is the following Akbota-II  equation 
 \begin{eqnarray}
 iq_t + \alpha q_{xx} +\beta q_{xt} +vq&=&0\label{2.28},\\
 v_{x}-2[\alpha(|q|^2)_{x}+\beta(|q|^2)_{t}] &=&0.\label{2.29}
\end{eqnarray}
The Lax representation of this Akbota-II  equation  is given by
\begin{eqnarray}
\Phi_{x}&=&U_{14}\Phi \label{2.31},\\
\Phi_{t}&=& V_{14}\Phi \label{2.31c},
\end{eqnarray}
where
\begin{eqnarray}
U_{14}=\frac{i\lambda}{2}\sigma_3+Q, \quad Q=\left(
\begin{array}{cc}
0 & \bar{q} \\
q & 0
\end{array}\right), \quad V_{14}=\frac{1}{1-\lambda\beta}\{\frac{i\lambda^2}{2}\alpha\sigma_3+\alpha\lambda Q+V_0\}\label{2.33c}
\end{eqnarray}
with
\begin{eqnarray}
 V_0=\left(
\begin{array}{cc}
\alpha i|q|^{2}+i\beta\partial_x^{-1}|q|^{2}_t & -i\beta  \bar{q}_{t}-i\alpha \bar{q}_{x} \\
i\beta q_y+\alpha i q_x & -[\alpha i|q|^{2}+i\beta\partial_x^{-1}|q|^{2}_{t}]
\end{array} \right) \label{2.34c}.
\end{eqnarray}

%%%%%%%%%%%%%%%%%%%%%%%%%%%%%%%%%%%%%%%%%%%%%%%%%%%%%%%%
\section{Integrable spin systems in 1+0 dimensions: Zhanbota transcendents and Zhanbota equations}
Interesting subclass of the integrable spin systems is  integrable Zhanbota equations in 1+1 and 1+0 dimensions. These  integrable Zhanbota equations are called as: \\
the Zhanbota-I equation,\\
the Zhanbota-II equation, \\ the Zhanbota-III equation, \\ the Zhanbota-IV equation,\\
the Zhanbota-V equation,\\ the Zhanbota-VI equation,\\ the Zhanbota-VII equation,  \\ the Zhanbota-VIII equation \\
etc. As integrable equations, these Zhanbota equations admit Lax representations, infinite number conservation laws, Hamiltonian structures  and so on. Among of these Zhanbota equations, six Zhanbota equations, namely, the Zhanbota-III, Zhanbota-IV, Zhanbota-V, Zhanbota-VI, Zhanbota-VII and Zhanbota-VIII equations  have the especial property. Zhanbota  transcendents are solutions of these six Zhanbota equations  in the complex plane. Note that these six Zhanbota equations  are not generally solvable in terms of elementary functions. It is interesting to note that these six Zhanbota equations correspond to the famous six Painleve equations. This remarkable property of these six Zhanbota equations we listed in the following Table 1:   
\\
\\
\\
\\
Table 1.  Some integrable Zhanbota equations and their gauge equivalent counterparts.
\\
\\
\begin{tabular} {|c|c|c|c|} \hline
No&Name&Equivalent counterpart&References\\ \hline

1&Zhanbota-III equation& Painleve-I equation
&[2205.02073]\\ \hline
2&Zhanbota-IV equation& 
Painleve-II equation&[2205.02073]\\ \hline
3&Zhanbota-V equation& 
Painleve-III equation &[2205.02073]\\ \hline
4&Zhanbota-VI equation& 
Painleve-IV equation&[2205.02073]\\ \hline
5&Zhanbota-VII equation& 
Painleve-V equation&[2205.02073]\\ \hline
6&Zhanbota-VIII equation& 
Painleve-VI equation
&[2205.02073]\\  \hline
\end{tabular}
\\
\\

Table 1. In this table we present (listed)  some integrable (1+0)-dimensional  Zhanbota equations (that is in 1+0 dimensions) as well as  their gauge  equivalent counterparts.  In this section, we want to present some of these six Zhanbota equations. 

\subsection{Zhanbota-III equation}

One of interesting integrable spin systems in 1+0 dimensions is the Zhanbota-III  equation. It has the form \cite{49}-\cite{2205.02073}
\begin{eqnarray}
S_{xx}+\frac{1}{2}tr(S_{x}^{2})S-\frac{1}{2}(\ln(tr(S_{x}^{2})))_{x}S_{x}+(\beta+\epsilon\sqrt{tr(S_{x}^{2}})[S,S_{x}]=0, 
\end{eqnarray}
where $\epsilon=\pm 1$ and $\beta$ is some complex constant. This Zhanbota-III  equation  is integrable. The corresponding Lax representation is given by
\begin{eqnarray}
\Psi_{x}&=&U_{III}\Psi, \\
\Psi_{\lambda}&=&V_{III}\Psi, 
\end{eqnarray}
where
\begin{eqnarray}
U_{III}&=&\left(\lambda-\beta+\frac{q_{III}}{\lambda}-\frac{q_{III}}{\beta}\right)S+\frac{\beta}{4}\left(\frac{1}{\lambda}-\frac{1}{\beta}\right)[S,S_{x}], \label{13.4}\\
V_{III}&=&(4\lambda^{4}+2q_{III}^{2}+x)S+\frac{\beta}{2q_{III}}\left(2\lambda q_{IIIx}+\frac{1}{2\lambda}\right)S_{x}+\frac{\beta}{4q_{III}}\left(4\lambda^{2}q_{III}+2q_{III}^{2}+x\right)[S,S_{x}],
\end{eqnarray}
with  
\begin{eqnarray}
q_{III}=\pm 0.5\beta\sqrt{{\bf S}_{x}^{2}}, \quad {\bf S}_{x}^{2}=0.5tr(S_{x}^{2}), \quad q_{III}^{2}=0.25\beta^{2}{\bf S}_{x}^{2}=\frac{\beta^{2}}{8}tr(S_{x}^{2}).\label{13.6}
\end{eqnarray}
The compatibility condition of the linear equations (6.2)-(6.3) $\Psi_{x\lambda}=\Psi_{\lambda x}$ that is
\begin{eqnarray}
U_{III\lambda}-V_{IIIx}+[U_{III},V_{III}]=0
\end{eqnarray}
  is equivalent to the Zhanbota-III equation (6.1). It is interesting to find the gauge equivalent counterpart of the Zhanbota-III equation. As it follows from the Table 1, the gauge equivalent counterpart of the Zhanbota-III equation is the Painleve-I equation. To prove it, consider the following well known representation of the spin matrix S:
	\begin{eqnarray}
S=\Phi^{-1}\sigma_{3}\Phi,
\end{eqnarray}
where $\Phi(x,\lambda)$ is a complex matrix. Substituting this formula to the Zhanbota-III equation (6.1), we obtain the following set of two linear equations
\begin{eqnarray}
\Phi_{x}&=&U_{III}^{\prime}\Phi, \\
\Phi_{\lambda}&=&V_{III}^{\prime}\Phi, 
\end{eqnarray}
where
\begin{eqnarray}
U_{III}^{\prime}&=&(\lambda+\frac{q_{III}}{\lambda})\sigma_{3}-\frac{iq_{III}}{\lambda}\sigma_{2}, \label{13.11}\\
V_{III}^{\prime}&=&(4\lambda^{4}+2q_{III}^{2}+x)\sigma_{3}-i(4\lambda^{2}q_{III}+2q_{III}^{2}+x)\sigma_{2}-(2\lambda q_{IIIx}+\frac{1}{2\lambda})\sigma_{1}.
\end{eqnarray}
From the compatibility condition of the two linear equations (6.9)-(6.10) that is from $\Phi_{x\lambda}=\Phi_{\lambda x}$, we obtain
\begin{eqnarray}
U_{III\lambda}^{\prime}-V^{\prime}_{IIIx}+[U^{\prime}_{III},V^{\prime}_{III}]=0.
\end{eqnarray}
This equation gives us  the following famous Painleve-I equation
\begin{eqnarray}
q_{IIIxx}-6q_{III}^{2}-x=0.
\end{eqnarray}
Thus we have proved that the Zhanbota-III equation (6.1) and and the Painleve-I equation (6.14) is gauge equivalent to each other. Note that takes place the following formulas:
\begin{eqnarray}
\Psi=g^{-1}\Phi, \quad g=\Phi|_{\lambda=\beta}.
\end{eqnarray}
\subsection{Zhanbota-IV equation}
Another example of integrable spin systems in 1+0 dimensions is the Zhanbota-IV   equation. The Zhanbota-IV equation   reads as \cite{49}-\cite{2205.02073}
\begin{eqnarray}
S_{xx}+\frac{1}{2}tr(S_{x}^{2})S-\frac{1}{2}\left(\ln\left(-\frac{tr(S_{x}^{2})}{8}\right)\right)_{x}S_{x}-i\beta[S,S_{x}]=0, \label{13.8}
\end{eqnarray}
where $\beta$ is an complex constant. As integrable equation, this Zhanbota-IV equation  admits the Lax representation, infinite number of conservation  laws and so on.  For example, its  Lax representation has the form
\begin{eqnarray}
\Psi_{x}&=&U_{IV}\Psi, \\
\Psi_{\lambda}&=&V_{IV}\Psi, 
\end{eqnarray}
where
\begin{eqnarray}
U_{IV}&=&-i(\lambda-\beta)S, \label{13.11}\\
V_{IV}&=&-i(4\lambda^{2}+2q_{IV}^{2}+x)S+\frac{iq_{IVx}}{q_{IV}}S_{x}+\left(\lambda-\frac{\alpha}{4\lambda q_{IV}}\right)[S,S_{x}].
\end{eqnarray}
In this case  
\begin{eqnarray}
q_{IV}=\pm i 0.25\sqrt{{\bf S}_{x}^{2}}, \quad {\bf S}_{x}^{2}=0.5tr(S_{x}^{2}), \quad q_{IV}^{2}=-0.25{\bf S}_{x}^{2}=-\frac{1}{8}tr(S_{x}^{2}).
\end{eqnarray}
From the compatibility condition of the linear equations (6.9)-(6.10) $\Psi_{x\lambda}=\Psi_{\lambda x}$:
\begin{eqnarray}
U_{IV\lambda}-V_{IVx}+[U_{IV},V_{IV}]=0
\end{eqnarray}
we obtain  the Zhanbota-IV equation (6.8). As in the previous subsection, we can prove that the gauge equivalent counterpart of the Zhanbota-IV equation (6.16) is the following Painleve-II equation:
\begin{eqnarray}
q_{IVxx}-2q_{IV}^{3}-xq_{IV}-\alpha=0.
\end{eqnarray}
\section{Other integrable equations}
There are some other integrable equations related with the Zhaidary equation. Let us present some of them. 
\subsection{The Kagaz-Ratbay-Myrzakul equation}
The Kagaz-Ratbay-Myrzakul equation (KRME) reads as
\begin{eqnarray}
iq_{t}+Aq_{xx}+Bq_{xy}-2A(vq)_{x}-\frac{1}{n}(b_{11}q)_{x}-2v[Bq_{y}+Aq_{x}-2Avq-\frac{1}{n}b_{11}q]-2ik_{11}q-i\alpha q_{y}&=&0, \label{7.1} \\
ir_{t}-Ar_{xx}-Br_{xy}-2A(vr)_{x}-\frac{1}{n}(b_{11}r)_{x}-2v[Br_{y}+Aq_{x}+2Avr+\frac{1}{n}b_{11}r]+2ik_{11}r-i\alpha r_{y}&=&0, \label{7.2}\\
v_{t}-k_{11x}-\alpha v_{y}&=&0, \label{7.3} \\
b_{11x}+iA(rq)_{x}+iB(rq)_{y}-2nBv_{y}&=&0, \label{7.4}
\end{eqnarray}
where $A, B, n, \alpha$  are some real constants. 
The  KRME    is integrable. Its   Lax representation is given by
\begin{eqnarray}
\Phi_{x}&=&U\Phi, \label{7.5}\\
\Phi_{t}&=&(\alpha-2nB\lambda^{2})\Phi_{y}+B\Phi, \label{7.6}
\end{eqnarray}
where
\begin{eqnarray}
U&=&(in\lambda^{2}+v)\sigma_{3}+\lambda Q,\\
B&=&\lambda^{4}B_{4}+\lambda^{3}B_{3}+\lambda^{2}B_{2}+\lambda B_{1}+B_{0},
\end{eqnarray}
and 
\begin{eqnarray}
\sigma_{3}&=&\begin{pmatrix} 1&0\\0 & -1\end{pmatrix}, \quad  Q=\begin{pmatrix} 0&q\\ r& 0\end{pmatrix}, \quad B_{4}=-2in^{2}A, \quad B_{3}=-2nAQ, \\
B_{2}&=&b_{11}\sigma_{3}, \quad B_{1}=\begin{pmatrix}0&c_{12}\\ 
c_{21}&0\end{pmatrix},  \quad B_{0}=k_{11}\sigma_{3},\quad \mu=-2nB,   \label{28}\\
c_{12}&=&-i[-B q_{y}-Aq_{x}+2Avq+\frac{1}{n}b_{11}q],\\
c_{21}&=&-i[B r_{y}+Ar_{x}+2Avr+\frac{1}{n}b_{11}r].  \label{27}
\end{eqnarray}
The compatibility condition $\Phi_{xt}=\Phi_{tx}$  of the  linear equations (\ref{7.5})-(\ref{7.6}) that is 
\begin{eqnarray}
U_{t}-B_{x}+[U,B]-(\alpha-2nB \lambda^{2}) U_{y}=0 \label{29} 
\end{eqnarray}
gives the KRME  (\ref{7.1})-(\ref{7.4})  \cite{2206.05348}. As the  integrable equation, the KRME  (\ref{7.1})-(\ref{7.4}) has  the  N-soliton solution, infinite number of conservation laws, Hamiltonian structure and so on.
 %%%%%%%%%%%%%%%%%%%%%%%%%%%%%%%%%%%%%%%%%%%%%%%%%%%%%%%%%%%
\subsection{The Kuralay-Ma-Myrzakulov equation}
%%%%%%%%%%%%%%%%%%%%%%%%%%%%%%%%%%%%%%%%%%%%%%%%%%%
One of interesting and important integrable equations in 2+1 dimensions is  the Kuralay-Ma-Myrzakulov equation (KMME). The KMME is given by
\begin{eqnarray}
Z_{t}-\frac{\beta }{b}f_{xy}Z+\frac{\beta}{b} f_{y}Z_{x}&=&0,\label{7.14} \\
f_{xxy}+\frac{b}{4a^{3}\beta}tr\left([Z_{xt},Z_{xx}]Z_{x}\right)&=&0,\label{7.15}
\end{eqnarray}
where $Z(x, y, t)$ is a $2\times 2$ matrix function, $f(x, y, t)$ is a  scalar function,  $(a,  b,  \beta)$ are some constants. Let us we introduce two new functions $q(x,y,t)$ and $u(x,y,t)$ as 
\begin{eqnarray}
q&=&2f_{x},\label{3} \\
u&=&-\beta f_{y}. \label{4}
\end{eqnarray}
Then, the KMME (\ref{7.14})-(\ref{7.15})  can  be rewritten in the equivalent form as
\begin{eqnarray}
Z_{t}-\frac{\beta}{2b} q_{y}Z-\frac{1}{b}uZ_{x}&=&0,\label{5} \\
q_{xy}+\frac{b}{2a^{3}\beta}tr\left([Z_{xt},Z_{xx}]Z_{x}\right)&=&0,\label{6}\\
u_{x}+0.5\beta q_{y}&=&0. \label{7}
\end{eqnarray}

This  KMME  is integrable. The corresponding Lax  representation is given by
\begin{eqnarray}
\Psi_{x}&=&U_{1}\Psi,\label{7.21}\\
\Psi_{t}&=&\beta\lambda\Psi_{y}+C\Psi,\label{7.22}
\end{eqnarray} 
where
\begin{eqnarray}
U_{1}&=&(b\lambda^{2}+q\lambda)Z,\label{13}\\
C&=&\lambda^{2}C_{2}+\lambda C_{1}.\label{14} 
\end{eqnarray}
Here
\begin{eqnarray}
C_{2}=uZ, \quad C_{1}=\frac{uq}{b}Z+\beta g^{-1}g_{y}.\label{15} 
\end{eqnarray}

\subsection{The Akbota--Myrzakulov-Tolkynay-Zhaidary equation}
%%%%%%%%%%%%%%%%%%%%%%%%%%%%%%%%%%%%%%%%%%%%%%%%%
Consider  the well-known Akbota--Myrzakulov-Tolkynay-Zhaidary  equation 
(AMTZE). The AMTZE  reads as
\begin{eqnarray}
2f_{xt}+\frac{2\beta}{b}f_{y}f_{xx}+\frac{4\beta}{b} f_{x}f_{xy}-\beta r_{y}&=&0, \label{18} \\
r_{t}+\frac{\beta}{b}f_{y}r_{x}+\frac{2\beta}{b} rf_{xy}-\frac{\beta}{2ab}f_{xxxy}&=&0.\label{19}
\end{eqnarray}
We can rewrite this AMTZE in the following equivalent form
\begin{eqnarray}
q_{t}-\frac{1}{b}uq_{x}+\frac{\beta}{b} qq_{y}-\beta r_{y}&=&0, \label{20} \\
r_{t}-\frac{1}{b}ur_{x}+\frac{\beta}{b} rq_{y}-\frac{\beta}{4ab}q_{xxy}&=&0,\label{21}\\
u_{x}+\frac{\beta}{2} q_{y}&=&0, \label{22}
\end{eqnarray}
where $a,b, \beta$ are real constants, $(q, r, u, f)$ are some functions of $(x,t,y)$. Note that the  AMTZE  is integrable. Its   Lax representation is given by  
\begin{eqnarray}
\Phi_{x}&=&U_{2}\Phi, \label{7.31}\\
\Phi_{t}&=&\beta\lambda\Phi_{y}+B\Phi, \label{7.32}
\end{eqnarray}
where
\begin{eqnarray}
U_{2}&=&\begin{pmatrix}0&a\\ b\lambda^{2}+q\lambda+r&0  \end{pmatrix}=(b\lambda^{2}+q\lambda)\Sigma+Q, \quad \sigma_{3}=\begin{pmatrix} 1&0\\0 & -1\end{pmatrix}. \label{25}  \\
B&=&B_{2}\lambda^{2}+B_{1}\lambda +B_{0}, \quad  \Sigma=\begin{pmatrix} 0&0\\1& 0\end{pmatrix}, \quad  Q=\begin{pmatrix} 0&a\\ r& 0\end{pmatrix}, \label{26}\\
B_{2}&=&\begin{pmatrix}0& 0 \\ u & 0
\end{pmatrix}=u\Sigma, \quad B_{1}=\begin{pmatrix}0& 0 \\ b^{-1}uq & 0
\end{pmatrix}=b^{-1}uq \Sigma, \label{27}\\
B_{0}&=&\begin{pmatrix}\frac{\beta }{4b}q_{y}& ab^{-1}u \\b^{-1}ur+\frac{\beta}{4ab}q_{xy}& -\frac{\beta }{4b}q_{y}
\end{pmatrix}=\frac{\beta }{4b}q_{y}\sigma_{3}+\frac{u}{b}Q+\frac{\beta}{4ab}q_{xy}\Sigma. \label{28}
\end{eqnarray}
The compatibility condition $\Phi_{xt}=\Phi_{tx}$  of the  linear equations (\ref{7.31})-(\ref{7.32}) that is 
\begin{eqnarray}
U_{2t}-B_{x}+[U_{2},B]-\beta \lambda U_{2y}=0 \label{29} 
\end{eqnarray}
gives the AMTZE. As the  integrable equation, the ATZME has  the  N-soliton solution, infinite number of conservation laws, Hamiltonian structure and so on. 

\subsection{The Zhaidary-Zhanbota-Myrzakulov equation}
The Zhaidary-Zhanbota-Myrzakulov equation reads as
 \begin{eqnarray}
iq_{t}+q_{xx}+(a_{0}+a_{1}|q|^{2}+a_{3}u+a_{4}u^{2})q&=&0, \\
b_{1}u_{xxx}-2b_{2}uu_{xx}+3b_{3}(u_{x})^{2}+\kappa(a_{0}+a_{1}|q|^{2}+a_{3}u+a_{4}u^{2})_{xx}&=&0.
\end{eqnarray}
\subsection{The Akbota-Kuralay-Myrzakulov  equation}
The Akbota-Kuralay-Myrzakulov  equation (AKME) reads as
\begin{eqnarray}
iq_{t}+Bq_{xt}-\frac{1}{n}(b_{11}q)_{x}-2v[Bq_{t}-\frac{1}{n}b_{11}q]-2ik_{11}q-i\alpha q_{t}&=&0, \label{20} \\
ir_{t}-Br_{xt}-\frac{1}{n}(b_{11}r)_{x}-2v[Br_{t}+\frac{1}{n}b_{11}r]+2ik_{11}r-i\alpha r_{t}&=&0, \label{21}\\
v_{t}-k_{11x}-\alpha v_{t}&=&0,  \\
b_{11x}+iB(rq)_{t}-2nBv_{t}&=&0, \label{22}
\end{eqnarray}
where $B, n, \alpha$ are constants. This AKME is integrable that is it has the Lax representation.  
\subsection{The Kairat-XII  equation}
Consider the following  Kairat-XII equation (K-XIIE)
\begin{eqnarray}
(1+6v)q_{t}-q^{3}q_{xxx}+3v_{t}q&=&0,\\
v_{x}-(q^{-1})_{t}&=&0.
\end{eqnarray}
This K-XIIE is integrable. The corresponding Lax representation is given by
\begin{eqnarray}
\psi_{xxx}+\frac{1}{4q}(1+6q_{x}-6v)\psi_{xx}&=&0,\\
\psi_{t}+q^{2}\psi_{xx}&=&0. 
\end{eqnarray}
This K-XIIE is integrable that is it has the Lax representation. The gauge equivalent counterpart of the K-XIIE is the Kairat-XI equation (K-XIE).
%%%%%%%%%%%%%%%%%%%%%%%%%%%%%%%%%%%%%%%%%%%%%%%%%
\subsection{The Aigerim-Ratbay-Shyngys equation}
The Aigerim-Ratbay-Shyngys equation (ARSE) is given by
\begin{eqnarray}
iq_{t}+q_{xt}+iq_{xxx}-uq-i(rq)q_{x}&=&0,\\
ir_{t}-r_{xt}+ir_{xxx}+ur-i(rq)q_{x}&=&0,\\
u_{x}+2k(rq)_{t}&=&0.
\end{eqnarray}
Here $q(x,t)$ and $r(x,t)$ are complex functions. $u(x.t)$ is a real function (potential), $k=const$. In the case, $r=\epsilon \bar{q}$ and $\epsilon=\pm 1$, this equation takes the form
\begin{eqnarray}
iq_{t}+q_{xt}+iq_{xxx}-uq-i(|q|^{2})q_{x}&=&0,\\
u_{x}+2k\epsilon(|q|^{2})_{t}&=&0.
\end{eqnarray}
Note that the ARSE is integrable. Its LR reads as
\begin{eqnarray}
\Phi_{x}&=&U\Phi,\\
\Phi_{t}&=&\frac{1}{1-2\lambda}B\Phi,
\end{eqnarray}
where
\begin{eqnarray}
U&=&a\lambda\sigma_{3}+Q, \\
B&=&\lambda^{3}B_{3}+\lambda^{2}B_{2}+
\lambda B_{1}+B_{0}.
\end{eqnarray}
and 
\begin{eqnarray}
B_{3}&=&4i\sigma_{3}, \quad B_{2}=4Q, 
\quad Q=\begin{pmatrix} 0&q\\ r & 0\end{pmatrix}, \\
B_{1}&=&2irq\sigma_{3}-2i\sigma_{3}Q_{x}, \quad \sigma_{3}=\begin{pmatrix} 1&0\\ 0& -1\end{pmatrix}, \\
B_{0}&=&h_{11}\sigma_{3}-Q_{xx}+2rqQ+i\sigma_{3}Q_{t}, \\
 h_{11}&=&-(r_{x}q-rq_{x})-0.5iu,
\quad u_{x}-2(rq)_{t}=0. \label{27}
\end{eqnarray}
The compatibility condition of the linear equations (7.53)-(7.54) gives the following form of the ARSE
\begin{eqnarray}
iq_{t}+i(q_{xxx}-6rqq_{x})+q_{xt}-uq&=&0,\\
ir_{t}+i(r_{xxx}-6rqr_{x})-(r_{xt}-ur)&=&0,\\
u_{x}-2(rq)_{t}&=&0.
\end{eqnarray}
After the reduction $r=\epsilon \bar{q}$, this ARSE takes the form
\begin{eqnarray}
iq_{t}+i(q_{xxx}-6\epsilon |q|^{2}q_{x})+q_{xt}-uq&=&0,\\
u_{x}-2\epsilon(|q|^{2})_{t}&=&0.
\end{eqnarray}
 Note that usually, this  ARSE we  write as
\begin{eqnarray}
iq_{t}+q_{xt}-uq+i(q_{xxx}-|q|^{2}q_{x})&=&0,\\
u_{x}+2\epsilon k(|q|^{2})_{t}&=&0,
\end{eqnarray}
where $k=const$.
\section{Conclusion}
%%%%%%%%%%%%%%%%%%%%%%%%%%%%%%%%%%%%%%%%%

In this paper, we have shown that the Zhaidary equation is integrable by the IST method. Its gauge equivalent equation is constructed. Some integrable particular cases (reductions) is studied. Also  several integrable generalizations of the ZE are presented.  For these reductions and generalizations their gauge partners and their Lax representations are found. Various issues are worthy of further exploration. In particular, we hope to investigate the discrete,  nonlocal and fractional integrable versions of the ZE, its reductions and generalizations as well as solutions and geometry  of the above presented integrable equations  in the future.  For example, the ZE and related integrable equations  admit a geometric interpretation
that directly relates the curvature and torsion of a vector field to any their solution. We will demonstrate that these geometrical relations and interpretations can be extended to the extended
 discrete,  nonlocal and fractional integrable versions of the ZE. Finally, we consider Zhanbota transcendents and  integrable Zhanbota equations. In particular, the gauge equivalence between some  Zhanbota equations and the famous  six Painleve equations is established.

\section{Acknowledgments} This work was supported  by  the Ministry of Education  and Science of Kazakhstan, Grant AP08857372.

\section{Appendix A. Integrable and nonintegrable Myrzakulov equations}
In this Appendix A, we present some Myrzakulov-N equations. Some of these ME are integrable, but others are nonintegrable. About our notations: below M-N equation means Myrzakulov-N equation, where N=0, I, II, III, IV, V, VI, ... .  For example, the M-IV equation means the Myrzakulov-IV equation.

\begin{tabular} {|c|c|c|c|} \hline
&Name&Equation&References\\ \hline

1&M-0 equation& 
$
\begin{aligned}
\textbf{S}_t=\sum^{n}_{j=1}\left\{\sum^{n}_{i=1}(\alpha^{2}_{ij}\textbf{S}\wedge \textbf{S}_{x_ix_j}+v_{ij}\textbf{S}_{x_i} \wedge\textbf{S}_{x_j})+\omega_j\textbf{S}_{x_j}\right\}+\omega_0\textbf{S},\\
u_{jx_{i}}-a^{2}_{ji} u_{ix_j}=2k_{ji}\textbf{S}(\textbf{S}_{x_i}\wedge\textbf{S}_{x_j})
\end{aligned}
$
&[1(p.504)]\\ \hline
2&M-I equation& 
$
\begin{aligned}
{\bf S}_{t}=({\bf S}\wedge {\bf S}_{y}+u{\bf S})_{x},\\
u_x=-{\bf S}\cdot({\bf S}_{x}\wedge {\bf S}_{y}) 
\end{aligned}
$
&[1(p.389)]\\ \hline
3&M-II equation& 
$
\begin{aligned}
\textbf{S}_t=(\textbf{S}\wedge\textbf{S}_y+u\textbf{S})_x+2cb^2\textbf{S}_y-4cv\textbf{S}_x,\\
u_x=-\textbf{S}(\textbf{S}_x\wedge\textbf{S}_y),\ \ \ \ v_x=\frac{1}{16b^2c^2}(\textbf{S}^{2}_{1x})_y
\end{aligned}
$
&[1(p.390)]\\ \hline
4&M-III equation& 
$
\begin{aligned}
\textbf{S}_t=(\textbf{S}\wedge\textbf{S}_y+u\textbf{S})_x+2b(cb+d)\textbf{S}_y-4cv\textbf{S}_x,\\
u_x=-\textbf{S}(\textbf{S}_x\wedge\textbf{S}_y), v_x=\frac{1}{4(2bc+d)^2}(\textbf{S}^{2}_{x})_y
\end{aligned}
$
&[1(p.391)]\\ \hline
5&M-IV equation& 
$
\begin{aligned}
\textbf{S}_t=-\frac{m}{4k^2}(\textbf{S}_{xy}+V_3\textbf{S})_x,\\
V_{1x}=-\frac{1}{2}(\textbf{S}^2_x)_y
\end{aligned}
$
&[1(p.442)]\\ \hline
6a&M-V equation& 
$
\begin{aligned}
iR_t=\frac{1}{2}[R,R_{xx}]+\frac{3}{2}[R^2,(R^2)_{xx}]
\end{aligned}
$
&[6(p.20)]\\ \hline
6b&M-V equation& 
$
\begin{aligned}
\hat{\gamma}_t=\frac{1}{2}[\hat{\gamma}_s,\hat{\gamma}_{ss}]+\frac{3}{2}[\hat{\gamma}^{2}_{s},(\hat{\gamma}^{2}_{s})_{ss}],\,\,\hat{\gamma}_{s}\in osp(2|1)
\end{aligned}
$
&[13(p.5)]\\ \hline
7&M-VI equation& 
$
\begin{aligned}

\end{aligned}
$
&[]\\ \hline
8&M-VII equation& 
$
\begin{aligned}

\end{aligned}
$
&[]\\ \hline
9&M-VIII equation& 
$
\begin{aligned}
\textbf{S}_t+\textbf{S}\wedge\textbf{S}_{\xi\xi}+w\textbf{S}_\xi=0, \\
w_\eta=\textbf{S}(\textbf{S}_\xi\wedge\textbf{S}_\eta)
\end{aligned}
$
&[1(p.392)]\\ \hline
10&M-IX equation& 
$
\begin{aligned}
{\bf S}_t = {\bf S} \wedge M_1{\bf S}+A_2{\bf S}_x+A_1{\bf S}_y,\\
M_2u=2\alpha^{2} {\bf S}({\bf S}_x \wedge {\bf S}_y)
\end{aligned}
$
&[1(p.394)]\\ \hline
11&M-X equation& 
$
\begin{aligned}
\textbf{S}_t=(3k^2+k_{xx}+3\alpha^2w)k^{-1}\textbf{S}_x,\\
w_{xx}=k_{yy}
\end{aligned}
$
&[1(p.445)]\\ \hline
12&M-XI equation& 
$
\begin{aligned}
\textbf{S}_t=\frac{w}{k}\textbf{S}_x, \\
w=\alpha q_{xxx}+\beta q_{yyy}-3\alpha(vq)_x-3\beta(wq)_y,\ \ \ v_y=k,\ \ \ w_x=m
\end{aligned}
$
&[1(p.445)]\\ \hline
13&M-XII equation& 
$
\begin{aligned}
u_t+2(uv)_x-\frac{4uvu_x}{1+u}=0, \\
v_t+2vv_x-\left(\frac{2uv^2}{1+u}\right)_x=0
\end{aligned}
$
&[15]\\ \hline
14&M-XIII equation& 
$
\begin{aligned}
iA_s=\frac{1}{2k}[A,A_s]_y+\frac{i}{k^2}(\rho A)_y
\end{aligned}
$
&[17]\\ \hline
15&M-XIV equation& 
$
\begin{aligned}
iA_t=i(fA)_x+\frac{1}{4\alpha}[A,A_t]_x+\frac{1}{\alpha}[A,W],\\
W_x=i(\alpha-\omega)[A,W]
\end{aligned}
$
&[arxiv:16]\\ \hline
16&M-XV equation& 
$
\begin{aligned}
\alpha^{2} S_{yy} - a(a+1) S_{xx} +
\{\alpha^{2} S^{2}_{y} - a(a+1) S^{2}_{x}\} S + A_{2}^{\prime \prime } SS_x + A_{1}^{\prime \prime } SS_y, \\
M_{2} u = \frac{\alpha^{2}}{2i} tr( S [S_{x}, S_{y}]) 
\end{aligned}
$
&[3(p.3)]\\ \hline
17&M-XVI equation& 
$
\begin{aligned}
\textbf{S}_t=\frac{n}{2k}(\textbf{S}\wedge\textbf{S}_y+u\textbf{S})-\frac{m}{4k^2}\left[\textbf{S}_{xy}+(\textbf{S}_x\textbf{S}_y-V_1)\textbf{S}-u\textbf{S}\wedge\textbf{S}_x\right]_x, \\
u_x=-\textbf{S}(\textbf{S}_x\wedge\textbf{S}_y), V_{1x}=-\frac{1}{2}(\textbf{S}^2_x)_y
\end{aligned}
$
&[1(p.468)]\\ \hline
\end{tabular}

\begin{tabular}{|c|c|c|c|} \hline
18&M-XVII equation& 
$
\begin{aligned}
\textbf{S}_t=\frac{1}{4}\textbf{S}_{xxx}-\frac{3}{4}\textbf{S}_{xyy}+C_1\textbf{S}_x+C_2\textbf{S}_y+C_3\textbf{S}, \\
V_x+iV_y=\frac{1}{8}\left[(\textbf{S}^2_x+\textbf{S}^2_y)_x-i(\textbf{S}^2_x+\textbf{S}^2_y)_y\right]
\end{aligned}
$
&[1(p.443)]\\ \hline
19&M-XVIII equation& 
$
\begin{aligned}
iS_t+\frac{1}{2}\left[S,S_{xx}+2\alpha(2b+1)S_{xy}+\alpha^2S_{yy}\right]+A^{'}_2S_x+A^{'}_1S_y=0, \\
\alpha^2u_{yy}-u_{xx}=\frac{\alpha^2}{2i}tr\left(S[S_x,S_y]\right)
\end{aligned}
$
&[4]\\ \hline
20&M-XIX equation& 
$
\begin{aligned}
iS_t+\frac{1}{2}\left[S,\alpha^2S_{yy}-4a(a+1)S_{xx}\right]+A^{"}_2S_x+A^{"}_1S_y=0, \\
M_2u=\frac{\alpha^2}{2i}tr\left(S[S_x,S_y]\right)
\end{aligned}
$
&[1(p.395)]\\ \hline
21&M-XX equation& 
$
\begin{aligned}
\textbf{S}_t+\textbf{S}\wedge\left\{(b+1)\textbf{S}_{\xi\xi}-b\textbf{S}_{\eta\eta}\right\}+bu_\eta\textbf{S}_\eta+(b+1)u_\xi\textbf{S}_\xi=0,  \\
u_{\xi\eta}=\textbf{S}(\textbf{S}_\xi\wedge\textbf{S}_\eta)
\end{aligned}
$
&[1(p.393)]\\ \hline
22&M-XXI equation& 
$
\begin{aligned}
k_{t} +k_{xxy}-2(kV_{1})_{x}+2V_{2}k=0,\,\,\, V_{1x}=Ek_{y},\,\,\, V_{2x}=-Ek_{xy}  
\end{aligned}
$
&[5]\\ \hline
23&M-XXII equation& 
$
\begin{aligned}
-iS_t=\frac{1}{2}\left([S,S_{y}]+2iuS\right)_x+\frac{i}{2}V_1S_x-2ia^2S_y, \\
u_x=-\textbf{S}(\textbf{S}_x\wedge\textbf{S}_y),\ \ \ V_{1x}=\frac{1}{4a^2}(\textbf{S}^2_x)_y
\end{aligned}
$
&[1(p.392)]\\ \hline
24&M-XXIII equation& 
$
\begin{aligned}
k_{xx}-\sigma^2k_{yy}+\frac{1}{2}R(x,y,t)e^{2k}+(2k_{tt}+3k^2_t)e^{2k}=0
\end{aligned}
$
&[5]\\ \hline
25&M-XXIV equation& 
$
\begin{aligned}
k_{xx}+\sigma^2k_{yy}+\frac{1}{4}R\sin k+\sin k(k_{tt}-\frac{1}{2}k_t^2)=0
\end{aligned}
$
&[5]\\ \hline
26&M-XXV equation& 
$
\begin{aligned}
(\frac{k_{tx}}{\cos k})_x\sin k-\sigma^2(\frac{k_{ty}}{\sin k}_y\cos k +
\frac{1}{2}(k_x^2)_t+ \frac{1}{2}(k_y^2)_t+(k_{xx}+ \sigma^2k_{yy})k_t=\\
=(\frac{1}{2}R+3)\sin k\cos kk_t
\end{aligned}
$
&[5]\\ \hline
27&M-XXVI equation& 
$
\begin{aligned}
(k_{xx} +k_{yy})_t +(k_{xx} +k_{yy})k_t=(- \frac {1}{2}R -3)e^{2k}k_t
\end{aligned}
$
&[5]\\ \hline
28&M-XXVII equation& 
$
\begin{aligned}
k_{xx} +k_{yy} - A(x,y)e^{-k} +(1+ \frac {R(x,y)}{3})e^{2k} =0
\end{aligned}
$
&[5]\\ \hline
29&M-XXVIII equation& 
$
\begin{aligned}
k_{xx} +k_{yy} +k_{tt} - \frac{R}{8} k^5 =0
\end{aligned}
$
&[5]\\ \hline
30&M-XXIX equation& 
$
\begin{aligned}
\textbf{S}_t=-\frac{m}{4a^2}\left(\textbf{S}_{xy}+V_3\textbf{S}-u\textbf{S}\wedge\textbf{S}_x\right)_x, \\
u_x=-\textbf{S}(\textbf{S}_x\wedge\textbf{S}_y),\,\,\, V_{1x}=-\frac{1}{2}(\textbf{S}^2_x)_y
\end{aligned}
$
&[1(p.462)]\\ \hline
31&M-XXX equation& 
$
\begin{aligned}
q_t=iq_{yx}-\frac{1}{2}[(V_1q)_x-qV_2-Eqq_y], \\
V_{1x}=Eq_y,\,\,\, V_{2x}=-Eq_{yx} 
\end{aligned}
$
&[5]\\ \hline      
32&M-XXXI equation& 
$
\begin{aligned}
bS_{\eta \eta} - (b+1) S_{\xi \xi}] +
\{bS_{\eta \eta} - (b+1) S_{\xi\xi}\} S + \\
+i(b+1)w_{\eta} SS_{\eta} + ibw_{\xi}SS_{\xi} =0, \\
w_{\xi \eta} = - \frac{1}{4i}tr(S[S_{\eta},S_{\xi}])
\end{aligned}
$
&[3(p.5)]\\ \hline    
33&M-XXXII equation& 
$
\begin{aligned}
q_{xt}-4aq+2p_x=0,\\
r_{xt}-4ar-2n_x,\\
a_x-0.5(rq)_t+qn-rp=0,\\
\eta_x+rp-qn=0,\\
p_x+2iwp+2\eta q=0,\\
n_x-2iwn-2\eta r=0
\end{aligned} $     
&[16]\\ \hline     
34&M-XXXIII equation& 
$
\begin{aligned}
2iS_t=[S,S_{xx}]+2iuS_x, \\
u_t+u_x+\alpha(u^2)_x+\beta u_{xxx}+\frac{\lambda}{4}tr(S^2_x)_x=0 
\end{aligned} $     
&[1(p.348)]\\ \hline   
35&M-XXXIV equation& 
$
\begin{aligned}
{\bf S}_t = {\bf S} \wedge {\bf S}_{xx}+u{\bf S}_{x},\\
u_{t}+4\eta({\bf S}^{2}_{x})_{x}=0
\end{aligned}$
&[1(p.484)]\\ \hline
36&M-XXXV equation& 
$
\begin{aligned}
2iS_t=[S,S_{xx}]+2iuS_x,\\
\rho u_{tt}=\nu^2_0u_{xx}+\alpha(u^2)_{xx}+\beta u_{xxxx}+\frac{\lambda}{4}tr(S^2_x)_x
\end{aligned}$
&[1(p.348)]\\ \hline
\end{tabular}

\begin{tabular}{|c|c|c|c|} \hline
37&M-XXXVI equation& 
$
\begin{aligned}
2iS_t=[S,S_{xx}]+2iuS_x,\\
\rho u_{tt}=\nu^2_0u_{xx}+\frac{\lambda}{4}tr(S^2_x)_x
\end{aligned}$
&[1(p.348)]\\ \hline
38&M-XXXVII equation& 
$
\begin{aligned}
2iS_t=[S,S_{xxxx}]+2\left\{((1+\mu)\vec{S}^2_x-u+m)[S,S_x]\right\}_x,\\
u_{t}+u_{x}+\alpha(u^2)_{x}+\beta u_{xxx}+\lambda(\vec{S}^2_x)_x=0
\end{aligned}$
&[1(p.348)]\\ \hline
39&M-XXXVIII equation& 
$
\begin{aligned}
2iS_t=[S,S_{xxxx}]+2\left\{((1+\mu)\vec{S}^2_x-u+m)[S,S_x]\right\}_x, \\ 
u_{t}+u_{x}+\lambda(\vec{S}^2_x)_x=0
\end{aligned}$
&[1(p.348)]\\ \hline
40&M-XXXIX equation& 
$
\begin{aligned}
2iS_t=[S,S_{xxxx}]+2\left\{((1+\mu)\vec{S}^2_x-u+m)[S,S_x]\right\}_x,\\
\rho u_{tt}=\nu^2_0u_{xx}+\alpha(u^2)_{xx}+\beta u_{xxxx}+\lambda(\vec{S}^2_x)_{xx}
\end{aligned}$
&[1(p.347)]\\ \hline
41&M-XXXX equation& 
$
\begin{aligned}
2iS_t=[S,S_{xxxx}]+2\left\{((1+\mu)\vec{S}^2_x-u+m)[S,S_x]\right\}_x,\\
\rho u_{tt}=\nu^2_0u_{xx}++\lambda(\vec{S}^2_x)_{xx}
\end{aligned}$
&[10(p.15)]\\ \hline
42&M-XXXXI equation& 
$
\begin{aligned}
2iS_t=\{(\mu \vec S^2_x - u +m)[S,S_x]\}_x,\\
u_t+u_x +\alpha(u^2)_x+\beta u_{xxx}+\lambda (\vec S^2_x)_{x} = 0
\end{aligned}$
&[10(p.15)]\\ \hline
43&M-XXXXII equation& 
$
\begin{aligned}
2iS_t=\{(\mu \vec S^2_x - u +m)[S,S_x]\}_x,\\
u_t+u_x +\lambda (\vec S^2_x)_x = 0
\end{aligned}$
&[10(p.15)]\\ \hline
44&M-XXXXIII equation& 
$
\begin{aligned}
2iS_t=\{(\mu \vec S^2_x - u +m)[S,S_x]\}_x,\\
\rho u _{tt}=\nu^2_0 u_{xx}+\alpha (u^2)_{xx}+\beta u_{xxxx}+ \lambda
(\vec S^2_x)_{xx}
\end{aligned}$
&[10(p.15)]\\ \hline
45&M-XXXXIV equation& 
$
\begin{aligned}
2iS_t=\{(\mu \vec S^2_x - u +m)[S,S_x]\}_x,\\
\rho u _{tt}=\nu^2_0 u_{xx}+\lambda(\vec S^2_x)_{xx}
\end{aligned}$
&[10(p.15)]\\ \hline
46&M-XXXXV equation& 
$
\begin{aligned}
2iS_t=[S,S_{xx}]+(uS_3+h)[S,\sigma_3],\\
u_t+u_x+\alpha(u^2)_x+\beta u_{xxx}+\lambda(S^2_3)_x=0
\end{aligned}$
&[10(p.14)]\\ \hline
47&M-XXXXVI equation& 
$
\begin{aligned}
2iS_t=[S,S_{xx}]+(uS_3+h)[S,\sigma_3],\\
u_t+u_x+\lambda(S^2_3)_x=0
\end{aligned}$
&[10(p.14)]\\ \hline
48&M-XXXXVII equation& 
$
\begin{aligned}
2iS_t=[S,S_{xx}]+(uS_3+h)[S,\sigma_3],\\
\rho u_{tt}=\nu^2_0 u_{xx}+\alpha(u^2)_{xx}+\beta u_{xxxx}+\lambda (S^2_3)_{xx}
\end{aligned}$
&[10(p.14)]\\ \hline
49&M-XXXXVIII equation& 
$
\begin{aligned}
2iS_t=[S,S_{xx}]+(uS_3+h)[S,\sigma_3],\\
\rho u_{tt}=\nu^2_0 u_{xx}+\lambda(S^2_3)_{xx}
\end{aligned}$
&[10(p.14)]\\ \hline
50&M-XXXXIX equation& 
$
\begin{aligned}
2iS_t=[S,S_{xx}]+(u+h)[S,\sigma_3],\\
u_t+u_x+\alpha(u^2)_x+\beta u_{xxx}+\lambda(S_3)_x=0
\end{aligned}$
&[10(p.14)]\\ \hline
51&M-XL equation& 
$
\begin{aligned}
2iS_t=[S,S_{xxxx}]+2\left\{((1+\mu)\vec{S}^2_x-u+m)[S,S_x]\right\}_x,\\
\rho u_{tt}=\nu^2_0u_{xx}+\lambda(\vec{S}^2_x)_{xx}
\end{aligned}$
&[1(p.347)]\\ \hline
52&M-XLI equation& 
$
\begin{aligned}
2iS_t=\left\{(\mu\vec{S}^2_x-u+m)[S,S_x]\right\}_x,\\
u_{t}+u_{x}+\alpha(u^2)_{x}+\beta u_{xxx}+\lambda(\vec{S}^2_x)_x=0
\end{aligned}$
&[1(p.347)]\\ \hline
53&M-XLII equation& 
$
\begin{aligned}
2iS_t=\left\{(\mu\vec{S}^2_x-u+m)[S,S_x]\right\}_x,\\
u_{t}+u_{x}+\lambda(\vec{S}^2_x)_x=0
\end{aligned}$
&[1(p.347)]\\ \hline
54&M-XLIII equation& 
$
\begin{aligned}
2iS_t=\left\{(\mu\vec{S}^2_x-u+m)[S,S_x]\right\}_x,\\
\rho u_{tt}=\nu^2_0u_{xx}+\alpha(u^2)_{xx}+\beta u_{xxxx}+\lambda(\vec{S}^2_x)_{xx}
\end{aligned}$
&[1(p.347)]\\ \hline
55&M-XLIV equation& 
$
\begin{aligned}
2iS_t=\left\{(\mu\vec{S}^2_x-u+m)[S,S_x]\right\}_x,\\
\rho u_{tt}=\nu^2_0u_{xx}+\lambda(\vec{S}^2_x)_{xx}
\end{aligned}$
&[1(p.347)]\\ \hline
56&M-XLV equation& 
$
\begin{aligned}
2iS_t=[S,S_{xx}]+(uS_3+h)[S,\sigma_3],\\
u_{t}+u_{x}+\alpha(u^2)_{x}+\beta u_{xxx}+\lambda({S}^2_3)_{x}
\end{aligned}$
&[1(p.347)]\\ \hline
\end{tabular}

\begin{tabular}{|c|c|c|c|} \hline
57&M-XLVI equation& 
$
\begin{aligned}
2iS_t=[S,S_{xx}]+(uS_3+h)[S,\sigma_3],\\
u_{t}+u_{x}+\lambda({S}^2_3)_{x}=0
\end{aligned}$
&[1(p.347)]\\ \hline
58&M-XLVII equation& 
$
\begin{aligned}
2iS_t=[S,S_{xx}]+(uS_3+h)[S,\sigma_3],\\
\rho u_{tt}=\nu^2_0u_{xx}+\alpha(u^2)_{xx}+\beta u_{xxxx}+\lambda({S}^2_3)_{xx}
\end{aligned}$
&[1(p.347)]\\ \hline
59&M-XLVIII equation& 
$
\begin{aligned}
2iS_t=[S,S_{xx}]+(uS_3+h)[S,\sigma_3],\\
\rho u_{tt}=\nu^2_0u_{xx}+\lambda({S}^2_3)_{xx}
\end{aligned}$
&[1(p.346)]\\ \hline
60&M-XLVIX equation& 
$
\begin{aligned}
2iS_t=[S,S_{xx}]+(u+h)[S,\sigma_3],\\
u_{t}+u_{x}+\alpha(u^2)_{x}+\beta u_{xxx}+\lambda({S}_3)_{x}=0
\end{aligned}$
&[1(p.346)]\\ \hline
61&M-L equation& 
$
\begin{aligned}
2iS_t=[S,S_{xx}]+(u+h)[S,\sigma_3],\\
u_{t}+u_{x}+\lambda({S}_3)_{x}=0
\end{aligned}$
&[1(p.346)]\\ \hline
62&M-LI equation& 
$
\begin{aligned}
2iS_t=[S,S_{xx}]+(u+h)[S,\sigma_3],\\
\rho u_{tt}=\nu^2_0u_{xx}+\alpha(u^2)_{xx}+\beta u_{xxxx}+\lambda({S}_3)_{xx}
\end{aligned}$
&[1(p.346)]\\ \hline
63&M-LII equation& 
$
\begin{aligned}
2iS_t=[S,S_{xx}]+(u+h)[S,\sigma_3],\\
\rho u_{tt}=\nu^2_0u_{xx}+\lambda({S}_3)_{xx}
\end{aligned}$
&[1(p.346)]\\ \hline
64&M-LIII equation& 
$
\begin{aligned}
iA_t+\frac{1}{2}[A,A_{xx}]+iu_1A_x+v_1[\sigma_3,A]=0
\end{aligned}$
&[18]\\ \hline
65&M-LIV equation& 
$
\begin{aligned}
2iS_t=n[S,S_{xxxx}]+2\left\{(\mu\vec{S}^2_x-u+m)[S,S_x]\right\}_x+h[S,\sigma_3]
\end{aligned}$
&[1(p.345)]\\ \hline
66&M-LV equation& 
$
\begin{aligned}
2iS_t=\left\{(\mu\vec{S}^2_x-u+m)[S,S_x]\right\}_x+h[S,\sigma_3]
\end{aligned}$
&[1(p.345)]\\ \hline
67&M-LVI equation& 
$
\begin{aligned}
2iS_t=[S,S_{xx}]+(uS_3+h)[S,\sigma_3]
\end{aligned}$
&[1(p.345)]\\ \hline
68&M-LVII equation& 
$
\begin{aligned}
2iS_t=[S,S_{xx}]+(u+h)[S,\sigma_3]
\end{aligned}$
&[1(p.345)]\\ \hline
69&M-LVIII equation& 
$
\begin{aligned}
u_{t}+2(uv)_{x}=0, \\
v_{t}+(v^{2}+w)_{x}=0, \\
w_{t}+\sigma w_{x}+2\delta u_{x}=0. 
\end{aligned}$
&[15]\\ \hline
70&M-LIX equation& 
$
\begin{aligned}
\alpha {\bf e}_{1y}=f_{1}{\bf e}_{1x}+
\sum_{j=1}^{n}b_{j}{\bf e}_{1}\wedge \frac{\partial^{j}}{\partial x^{j}}{\bf e}_{1} +
c_{1}{\bf e}_{2}+d_{1}{\bf e}_{3} 
\end{aligned}$
&[6(p.3)]\\ \hline
71&M-LX equation& 
$
\begin{aligned}
\alpha\left ( \begin{array}{ccc}
{\bf e}_{1} \\
{\bf e}_{2} \\
{\bf e}_{3}
\end{array} \right)_{y}= A
\left ( \begin{array}{ccc}
{\bf e}_{1} \\
{\bf e}_{2} \\
{\bf e}_{3}
\end{array} \right)_{x}  + B
\left ( \begin{array}{ccc}
{\bf e}_{1} \\
{\bf e}_{2} \\
{\bf e}_{3}
\end{array} \right),\\
\left ( \begin{array}{ccc}
{\bf e}_{1} \\
{\bf e}_{2} \\
{\bf e}_{3}
\end{array} \right)_{t}=
\sum_{j=0}^{n}C_{j}\frac{\partial^{j}}{\partial x^{j}}
\left ( \begin{array}{ccc}
{\bf e}_{1} \\
{\bf e}_{2} \\
{\bf e}_{3}
\end{array} \right)
\end{aligned}$
&[1(p.453)]\\ \hline
72&M-LXI equation& 
$
\begin{aligned}
u_t+2(uv)_x+\frac{3}{2}\delta(u^2)_x=0,\\
v_{t}+(v^{2}+\delta uv)_{x}=0
\end{aligned}
$
&[15]\\ \hline

73&M-LXII equation& 
$
\begin{aligned}
u_{t}+2(uv)_{x}+\frac{1}{2}\delta (u^{2})_{x}=0, \\
v_{t}+(v^{2}+\delta uv)_{x}=0. 
\end{aligned}$
&[15]\\ \hline
74&M-LXIII equation& 
$
\begin{aligned}
u_{t}+2v_{y}u+vu_{y}+u_{x}\partial_{x}^{-1}v_{y}=0,\\
v_{t}+(v\partial_{x}^{-1}v_{y})_{x}+\omega_x=0,\\
\omega_x+2\delta u_{y}=0, 
\end{aligned}$
&[15]\\ \hline
75&M-LXIV equation& 
$
\begin{aligned}
iS_{t}+\epsilon_2i[ S_{xxx}+6(\alpha^2  S)_{x}]+\frac{1}{\omega}[S, W]=0,\\
iW_{x}+\omega [S, W]=0
\end{aligned}$
&[7(p.362)]\\ \hline
\end{tabular}
 
\begin{tabular}{|c|c|c|c|} \hline
76&M-LXV equation& 
$
\begin{aligned}
\hat e_{1x} = 2q\hat e_{2}-2p\hat e_{3} +\beta \hat e_{4}
-\epsilon\hat e_{5},\\
\hat e_{2x} = p e_{1}-2i\lambda \hat e_{2}+\epsilon\hat e_{4},\\
\hat e_{3x} = -q e_{1}+2i\lambda \hat e_{3}+\beta\hat e_{5},\\
\hat e_{4x} = \epsilon e_{1}-2\beta \hat e_{2} -i\lambda \hat e_{4}
-p\hat e_{5},\\
\hat e_{5x} =-\beta e_{1}+2\epsilon \hat e_{2} -q\hat e_{4} +
i\lambda \hat e_{5} 
\end{aligned}$
&[6(p.19)]\\ \hline
77a &M-LXVI equation& 
$
\begin{aligned}
k^{2} + \tau^{2} + \sigma^{2} =  n^{2}(x,y,z,t)
\end{aligned}$
&[8(p.15)]\\ \hline
77b &M-LXVI equation& 
$
\begin{aligned}
\left( \begin{array}{c}
{\bf e}^{\prime}_{1}\\
{\bf e}^{\prime}_{2}  \\
{\bf e}^{\prime}_{3}
\end{array} \right)_{x} = -\lambda A_{3}
\left( \begin{array}{c}
{\bf e}^{\prime}_{1}\\
{\bf e}^{\prime}_{2}  \\
{\bf e}^{\prime}_{3}
\end{array} \right),\\
\left( \begin{array}{c}
{\bf e}^{\prime}_{1}\\
{\bf e}^{\prime}_{2}  \\
{\bf e}^{\prime}_{3}
\end{array} \right)_{t} = \lambda
\left( \begin{array}{c}
{\bf e}^{\prime}_{1}\\
{\bf e}^{\prime}_{2}  \\
{\bf e}^{\prime}_{3}
\end{array} \right)_{y} -\lambda  A_{4}
\left( \begin{array}{c}
{\bf e}^{\prime}_{1}\\
{\bf e}^{\prime}_{2}  \\
{\bf e}^{\prime}_{3}
\end{array} \right)
\end{aligned}
$
&[14(p.9)]\\ \hline
77c &M-LXVI equation& 
$
\begin{aligned}
\left( \begin{array}{c}
{\bf e}_{1}\\
{\bf e}_{2}  \\
{\bf e}_{3}
\end{array} \right)_{\xi_1} = A
\left( \begin{array}{c}
{\bf e}_{1}\\
{\bf e}_{2}  \\
{\bf e}_{3}
\end{array} \right)_{\xi_3} + B
\left( \begin{array}{c}
{\bf e}_{1}\\
{\bf e}_{2}  \\
{\bf e}_{3}
\end{array} \right),\\
\left( \begin{array}{c}
{\bf e}_{1}\\
{\bf e}_{2}  \\
{\bf e}_{3}
\end{array} \right)_{\xi_2} = C
\left( \begin{array}{c}
{\bf e}_{1}\\
{\bf e}_{2}  \\
{\bf e}_{3}
\end{array} \right)_{\xi_4} + D
\left( \begin{array}{c}
{\bf e}_{1}\\
{\bf e}_{2}  \\
{\bf e}_{3}
\end{array} \right)
\end{aligned}
$
&[1(p.552)]\\ \hline
78a&M-LXVII equation& 
$
\begin{aligned}
\begin{matrix}\quad\left(\begin{array}{cc}
\textbf{e}_1\\
\textbf{e}_2\\
\textbf{e}_3\\
\vdots\\
\textbf{e}_n
\end{array}\right)_{\xi_1}
\end{matrix}=B\begin{matrix}\quad\left(\begin{array}{cc}
\textbf{e}_1\\
\textbf{e}_2\\
\textbf{e}_3\\
\vdots\\
\textbf{e}_n
\end{array}\right)
\end{matrix},\\
\begin{matrix}\quad\left(\begin{array}{cc}
\textbf{e}_1\\
\textbf{e}_2\\
\textbf{e}_3\\
\vdots\\
\textbf{e}_n
\end{array}\right)_{\xi_2}
\end{matrix}=C\begin{matrix}\quad\left(\begin{array}{cc}
\textbf{e}_1\\
\textbf{e}_2\\
\textbf{e}_3\\
\vdots\\
\textbf{e}_n
\end{array}\right)_{\xi_4}
\end{matrix}+D 
\begin{matrix}\quad\left(\begin{array}{cc}
\textbf{e}_1\\
\textbf{e}_2\\
\textbf{e}_3\\
\vdots\\
\textbf{e}_n
\end{array}\right)
\end{matrix}
\end{aligned}$
&[1(p.552)]\\ \hline

78b&M-LXVII equation& 
$
\begin{aligned}
\begin{matrix}\quad\left(\begin{array}{cc}
\textbf{e}_1\\
\textbf{e}_2\\
\textbf{e}_3\\
\vdots\\
\textbf{e}_n
\end{array}\right)_{\xi_1}
\end{matrix}=(A_1-\lambda A_3)\begin{matrix}\quad\left(\begin{array}{cc}
\textbf{e}_1\\
\textbf{e}_2\\
\textbf{e}_3\\
\vdots\\
\textbf{e}_n
\end{array}\right)
\end{matrix},\\
\begin{matrix}\quad\left(\begin{array}{cc}
\textbf{e}_1\\
\textbf{e}_2\\
\textbf{e}_3\\
\vdots\\
\textbf{e}_n
\end{array}\right)_{\xi_2}
\end{matrix}=\lambda\begin{matrix}\quad\left(\begin{array}{cc}
\textbf{e}_1\\
\textbf{e}_2\\
\textbf{e}_3\\
\vdots\\
\textbf{e}_n
\end{array}\right)_{\xi_4}
\end{matrix}+(A_2-\lambda A_4) 
\begin{matrix}\quad\left(\begin{array}{cc}
\textbf{e}_1\\
\textbf{e}_2\\
\textbf{e}_3\\
\vdots\\
\textbf{e}_n
\end{array}\right)
\end{matrix}
\end{aligned}$
&[1(p.552)]\\ \hline
\end{tabular}
 
\begin{tabular}{|c|c|c|c|} \hline
79&M-LXVIII equation& 
$
\begin{aligned}
\begin{matrix}\quad\left(\begin{array}{cc}
\textbf{e}_1\\
\textbf{e}_2\\
\textbf{e}_3\\
\vdots\\
\textbf{e}_n
\end{array}\right)_{\xi_1}
\end{matrix}=A\begin{matrix}\quad\left(\begin{array}{cc}
\textbf{e}_1\\
\textbf{e}_2\\
\textbf{e}_3\\
\vdots\\
\textbf{e}_n
\end{array}\right)_{\xi_3}
\end{matrix}+B\begin{matrix}\quad\left(\begin{array}{cc}
\textbf{e}_1\\
\textbf{e}_2\\
\textbf{e}_3\\
\vdots\\
\textbf{e}_n
\end{array}\right)\end{matrix},\\
\begin{matrix}\quad\left(\begin{array}{cc}
\textbf{e}_1\\
\textbf{e}_2\\
\textbf{e}_3\\
\vdots\\
\textbf{e}_n
\end{array}\right)_{\xi_2}
\end{matrix}=C\begin{matrix}\quad\left(\begin{array}{cc}
\textbf{e}_1\\
\textbf{e}_2\\
\textbf{e}_3\\
\vdots\\
\textbf{e}_n
\end{array}\right)_{\xi_4}
\end{matrix}+D\begin{matrix}\quad\left(\begin{array}{cc}
\textbf{e}_1\\
\textbf{e}_2\\
\textbf{e}_3\\
\vdots\\
\textbf{e}_n
\end{array}\right)\end{matrix}

\end{aligned}
$
&[1(p.547)]\\ \hline
80&M-LXIX equation& 
$
\begin{aligned}
\textbf{S}_t=\frac{1}{\sqrt{\textbf{S}^2_x}}\left(-\sqrt{\textbf{S}^2_x-u^2}\textbf{S}_x+u\textbf{S}\wedge\textbf{S}_x\right),\\
u_x=v\sqrt{\textbf{S}^2_t-u^2},\\
v_t=-\textbf{S}\cdot(\textbf{S}_t\wedge\textbf{S}_x)
\end{aligned}
$
&[1(p.485)]\\ \hline
81&M-LXX equation& 
$
\begin{aligned}
u_t+2\sqrt{1-u}(uv)_x=0,\\
v_t+(v^2\sqrt{1-u})_x+j_{31}(\sqrt{1-u})_x=0
\end{aligned}
$
&[15]\\ \hline
82&M-LXXI equation& 
$
\begin{aligned}
u_t+2(uv)_x-\frac{4uvu_x}{1+u}=0,\\
v_t+2vv_x-\left(\frac{2uv^2}{1+u}\right)_x+\frac{2j_{31}u}{1+u}=0
\end{aligned}
$
&[15]\\ \hline
83&M-LXXII equation& 
$
\begin{aligned}
iS_t+\frac{1}{4\lambda_0^2}[S,S_{xx}-S_{xt}]+\frac{i(1-4\lambda_0^4)}{4\lambda_0^4}S_x=0
\end{aligned}
$
&[?]\\ \hline
84&M-LXXIII equation& 
$

$$ 
[S_t,S_x]+[S,S_{xt}]+4\lambda_0 S_t+(2\lambda_0 |q|^{2} S+|q|^{2}SS_x-\frac{1}{\lambda_0}S)_x=0
$$

$
&[?]\\ \hline
85&M-LXXIV equation& 
$
\begin{aligned}
u_t+2\sqrt{1-u}(uv)_x=0,\\
v_t+(v^2\sqrt{1-u})_x=0
\end{aligned}
$
&[15]\\ \hline
86a&M-LXXV equation& 
$
\begin{aligned}
L_2\psi=\alpha\psi_y+\psi_{xx}+\beta (u\psi)_x=0,\\
\psi_t=A_1\psi
\end{aligned}
$
&[1(p.317)]\\ \hline
86b&M-LXXV equation& 
$
\begin{aligned}
\textbf{S}_t=\left\{(\mu\left|\phi\right|^2+m)\textbf{S}\wedge\textbf{S}_x\right\}_x+\textbf{S}\wedge\textbf{J}\textbf{S},\\
i\phi_x+\phi_{xx}+i\lambda (\textbf{S}^{2}_{x}\phi)_x=0\end{aligned}
$
&[11(p.541)]\\ \hline
87&M-LXXVI equation& 
$
\begin{aligned}
i\phi_t+\phi_{xx}+i(u\phi)_x=0,\\
\rho u_{tt}=\upsilon^2_0u_{xx}+\lambda(\left|\phi\right|^2)_{xx}
\end{aligned}
$
&[1(p.315)]\\ \hline
88&M-LXXVII equation& 
$
\begin{aligned}
i\phi_t+\phi_{xx}+i(u\phi)_x=0,\\
\rho u_{tt}=\upsilon^2_0u_{xx}+\alpha(u^2)_{xx}+\beta u_{xxxx}+\lambda(\left|\phi\right|^2)_{xx}
\end{aligned}
$
&[1(p.315)]\\ \hline
89&M-LXXVIII equation& 
$
\begin{aligned}
i\phi_t+\phi_{xx}+i(u\phi)_x=0,\\
u_{t}+u_{x}+\lambda(\left|\phi\right|^2)_{x}=0
\end{aligned}
$
&[1(p.315)]\\ \hline

90&M-LXXIX equation& 
$
\begin{aligned}
i\phi_t+\phi_{xx}+i(u\phi)_x=0,\\
u_{t}+u_{x}+\alpha(u^2)_{x}+\beta u_{xxx}+\lambda(\left|\phi\right|^2)_{x}=0
\end{aligned}
$
&[1(p.315)]\\ \hline
91&M-LXXX equation& 
$
\begin{aligned}
i\phi_t+\left[\frac{\phi}{\sqrt{1+u}}\right]_{xx}=0,\\
\rho u_{tt}=\nu^2_0u_{xx}+\lambda(\left|\phi\right|^2)_{xx}=0
\end{aligned}
$
&[1(p.315)]\\ \hline
92&M-LXXXI equation& 
$
\begin{aligned}
i\phi_t+\left[\frac{\phi}{\sqrt{1+u}}\right]_{xx}=0,\\
\rho u_{tt}=\nu^2_0u_{xx}+\alpha(u^2)_{xx}+\beta u_{xxxx}+\lambda(\left|\phi\right|^2)_{xx}
\end{aligned}
$
&[1(p.315)]\\ \hline
\end{tabular}

\begin{tabular}{|c|c|c|c|} \hline
93&M-LXXXII equation& 
$
\begin{aligned}
i\phi_t+\left[\frac{\phi}{\sqrt{1+u}}\right]_{xx}=0,\\
u_{t}+u_{x}+\lambda(\left|\phi\right|^2)_{x}=0
\end{aligned}
$
&[1(p.315)]\\ \hline
94&M-LXXXIII equation& 
$
\begin{aligned}
i\phi_t+\left[\frac{\phi}{\sqrt{1+u}}\right]_{xx}=0,\\
u_{t}+u_{x}+\alpha(u^2)_{x}+\beta u_{xxx}+\lambda(\left|\phi\right|^2)_{x}=0
\end{aligned}
$
&[1(p.315)]\\ \hline
95&M-LXXXIV equation& 
$
\begin{aligned}
i\phi_x+\psi+u\phi=0,\,\,i\psi_t+\phi+u\psi=0,\\ \rho u_{tt}=\nu^2_0u_{xx}+\lambda(\phi\psi)_{xx}

\end{aligned}
$
&[1(p.315)]\\ \hline
96&M-LXXXV equation& 
$
\begin{aligned}
i\phi_x+\psi+u\phi=0,\,\, i\psi_t+\phi+u\psi=0,\\ \rho u_{tt}=\nu^2_0u_{xx}+\alpha(u^2)_{xx}+\beta u_{xxxx}+\lambda(\phi\psi)_{xx}

\end{aligned}
$
&[1(p.316)]\\ \hline
97&M-LXXXVI equation& 
$
\begin{aligned}
i\phi_x+\psi+u\phi=0,\\
i\psi_t+\phi+u\psi=0, \\
u_{t}+u_{x}+\lambda(\phi\psi)_{x}=0
\end{aligned}
$
&[1(p.316)]\\ \hline
98&M-LXXXVII equation& 
$
\begin{aligned}
i\phi_x+\psi+u\phi=0,\\
i\psi_t+\phi+u\psi=0, \\
u_{t}+u_{x}+\alpha(u^2)_{x}+\beta u_{xxx}+\lambda(\phi\psi)_{x}=0
\end{aligned}
$
&[1(p.316)]\\ \hline
99&M-LXXXVIII equation& 
$
\begin{aligned}
i\phi_t+\phi_{xx}+iu\phi_x=0,\\
\rho u_{tt}=\nu^2_0u_{xx}+\lambda({\left|\phi\right|}^2)_{xx}
\end{aligned}
$
&[1(p.316)]\\ \hline
100&M-LXXXIX equation& 
$
\begin{aligned}
i\phi_t+\phi_{xx}+iu\phi_x=0,\\
\rho u_{tt}=\nu^2_0u_{xx}+\alpha(u^2)_{xx}+\beta u_{xxxx}+\lambda({\left|\phi\right|}^2)_{xx}
\end{aligned}
$
&[1(p.316)]\\ \hline
101&M-XC equation& 
$
\begin{aligned}
i\phi_t+\phi_{xx}+iu\phi_x=0,\\
u_{t}+u_{x}+\lambda({\left|\phi\right|}^2)_{x}
\end{aligned}
$
&[1(p.316)]\\ \hline
102&M-XCI equation& 
$
\begin{aligned}
i\phi_t+\phi_{xx}+iu\phi_x=0,\\
u_{t}+u_{x}+\alpha(u^2)_x+\beta u_{xxx}+\lambda({\left|\phi\right|}^2)_{x}=0
\end{aligned}
$
&[1(p.316)]\\ \hline
103&M-XCII equation& 
$
\begin{aligned}
{\bf \gamma}_{st}=(\alpha {\bf \gamma}_{ss}^{2}+\beta u +\delta){\bf \gamma}_{ss}
\end{aligned}
$
&[13(p.3)]\\ \hline
104&M-XCIII equation& 
$
\begin{aligned}
{\bf \gamma}_{st}=(\alpha \sqrt{{\bf \gamma}_{ss}^{2}}+\beta u +\delta){\bf \gamma}_{ss} 
\end{aligned}
$
&[13(p.3)]\\ \hline

105&M-XCIV equation& 
$
\begin{aligned}
iS_{t}+0.25\epsilon_1[S, S_{xx}]+\epsilon_2i[ S_{xxx}+6(\beta^2  S)_{x}]+\frac{1}{\omega}[S, W]=0,\\
iW_{x}+\omega [S, W]=0 
\end{aligned}
$
&[7(p.363)]\\ \hline
106&M-XCV equation& 
$
\begin{aligned}
\gamma_{st}=(\frac{3}{4}V-\frac{1}{2}\gamma^2_{ss}+W)\gamma_{ss},\\
V_s=(\sqrt{\gamma^2_{ss}})_y,\\
(W\sqrt{\gamma^2_{ss}})_s=(\frac{3}{4}V_y-\frac{3}{2}\gamma^2_{ss})_y
\end{aligned}
$
&[13(p.8)]\\ \hline

107&M-XCVI equation& 
$
\begin{aligned}
{\bf \gamma}_{st}=[W-3\partial^{-1}_{\bar z}(\sqrt{{\bf \gamma}^2_{sz}})_z]{\bf\gamma}_{sz},\\
W_z=-3[\sqrt{{\bf\gamma}^2_{sz}}\partial^{-1}_z(\sqrt{{\bf\gamma}^2_{sz}})_{\bar z}]_{\bar z}
\end{aligned}
$
&[13(p.8)]\\ \hline
108&M-XCVII equation& 
$
\begin{aligned}
(\gamma_{st}-\frac{3}{4}\sqrt{\gamma^2_{ss}}\gamma_{ss})_s=-\frac{3}{4i}
(\gamma_{syy}\cdot
\sigma_2\gamma_s)
\end{aligned}
$
&[13(p.8)]\\ \hline

109&M-XCVIII equation& 
$
\begin{aligned}
\gamma_{st}=\frac{3}{4}\sqrt{\gamma^2_{ss}}\gamma_{ss}
\end{aligned}
$
&[13(p.8)]\\ \hline

110&M-XCIX equation& 
$
\begin{aligned}
iS_{t}+0.25\epsilon_1[S, S_{xx}]+\frac{1}{\omega}[S, W]=0,\\
iW_{x}+\omega [S, W]=0
\end{aligned}
$
&[7 (p.362)]\\ \hline
111&M-C equation& 
$
\begin{aligned}
\gamma_{st}=f_1\gamma_s\times\gamma_{ss}+f_2\gamma_{ss}+f_3\gamma\times\gamma_{s}
\end{aligned}
$
&[13 (p.8)]\\ \hline
112&M-CXI equation& 
$
\begin{aligned}
\left[A,A_{xt}\right]+(\phi[A,A_{x}])_{x}+\frac{4}{\beta^{2}}A_{x}=0
\end{aligned}
$
&[19]\\ \hline

113&M-CXII equation& 
$
\begin{aligned}
{\bf A}_{xt}+\phi {\bf A}_{xx}+v_{1}{\bf A}+v_{2}{\bf A}_{x}+\frac{\alpha_{0}}{\beta^{2}}{\bf A}\wedge {\bf A}_{x}=0
\end{aligned}
$
&[19]\\ \hline

\end{tabular}

\section{Appendix B. Integrable Myrzakulov-Lakshmanan equations}
In this Appendix B, we present some  Myrzakulov-Lakshmanan equations (MLE). These equations are integrable.

\begin{tabular} {|c|c|c|c|} \hline
&Name&Equation&References\\ \hline
1&ML-I equation& 
$
\begin{aligned}
\textbf{S}_{t}-\textbf{S}\wedge(\alpha S_{xx}+\beta S_{xy})-u\textbf{S}_x=0, \\
u_x+\textbf{S}\times(\textbf{S}_x\wedge S_y)=0
\end{aligned}
$
&[12(p.1356)]\\ \hline
2&ML-II equation& 
$
\begin{aligned}
iS_{t}+\frac{1}{2}[S, S_{xy}]+iuS_{x}+\frac{1}{\omega}[S, W]=0,\\
u_x-\frac{i}{4}tr(S[S_x,S_y])=0,\\
 iW_{x}+\omega [S, W]=0
\end{aligned}
$
&[9(p.6)]\\ \hline
3&ML-III equation& 
$
\begin{aligned}
iS_{t}+i\epsilon_2(S_{xy}+[S_x,Z])_{x}+(wS)_{x}+\frac{1}{\omega}[S, W]=0,\\
u_x-\frac{i}{4}tr(S\times[S_x,S_y])=0,\\
w_x-\frac{i}{4}\epsilon_2[tr(S_x^2)]_y=0,\\
iW_{x}+\omega [S, W]=0
\end{aligned}
$
&[9(p.9)]\\ \hline
4&ML-IV equation& 
$
\begin{aligned}
iS_{t}+2\epsilon_1 Z_x+i\epsilon_2(S_{xy}+[S_x,Z])_{x}+(wS)_{x}+\frac{1}{\omega}[S, W]=0,\\
u_x-\frac{i}{4}tr(S\times[S_x,S_y])=0,\\
w_x-\frac{i}{4}\epsilon_2[tr(S_x^2)]_y=0,\\
iW_{x}+\omega [S, W]=0
\end{aligned}
$
&[12(p.1365)]\\ \hline
\end{tabular}

\end{document}